\documentclass[a4paper,12pt]{article}
\usepackage{amsmath,amssymb}
\usepackage[top=3cm,bottom=3cm,left=2cm,right=2cm]{geometry}
\usepackage{color}
\usepackage[T1]{fontenc}
\usepackage{hyperref}
\hypersetup{colorlinks=true}
\usepackage{cite}
\date{}
\begin{document}
\author{Haci  Akbas ${^\dag}$ and O. Teoman Turgut$^\ddag$ \\ Department of Physics, Bo\u{g}azi\c{c}i University \\ 34342 Bebek, Istanbul, Turkey \\ $^\dag$akbas@gmail.com, $^\ddag$turgutte@boun.edu.tr}
\title{\bf Born-Oppenheimer Approximation for a Simple Renormalizable System}
\maketitle
\begin{abstract}
We discuss a  simple singular system in two dimension, two heavy particles interacting with a light particle via an attractive contact interaction. Although intuitively clear the actual application of the  Born-Oppenheimer approximation to this problem is quite subtle. Nevertheless, with due care we show that this can be done, as a result we calculate the leading term of the Born-Oppenheimer approximation and indicate  how to get the higher order corrections.
\end{abstract}
\section{Introduction}
Born-Oppenheimer approximation \cite{Born-Opp} is the basic tool  of molecular physics \cite{landau, bethe-jackiw,weinbergQM}. When  all the particles interact via  Coulomb forces, as is the typical case of molecular physics, the Born-Oppenheimer vibrational energy levels go with $(m/M)^{1/2}$, where $m/M$ refers to the  light mass to heavy mass ratio. One can consider rotational energy levels as well as anharmonic vibrations as higher order corrections, since they turn out to be of order $(m/M)$. In the usual approach the relevant expansion parameter is thought to be  $(m/M)^{1/4}$, yet only even powers seem to show up in the expansions. There is a large literature on the stationary level calculations in the Born-Oppenheimer approximation, we will not be able to cover most of it, we only mention some of the more rigorous works, since in our present work, we also aim to get an approximation scheme in which we can control the errors that we make in a self-consistent way. The time-dependent Born-Oppenheimer approximation is a very interesting and closely related subject,  the reader can consult the review articles \cite{Spohn, Hagedorn-Joye, Jecko-5} for more information. Albeit, our system is highly singular and requires renormalization as is well known in the literature\cite{Thorn, Beg, Jackiw, Mead, Perez, Albeverio, Mitra, Tarrach}. Therefore, one does not expect the existing results on well-behaved potentials to hold in this regime.

An interesting toy model in which two heavy and one light particle all interact via harmonic oscillator potentials is presented by R. Seiler in \cite{seiler-2}, where the assumptions of the Born-Oppenheimer approach is carefully tested. Following this, some  rigorous aspects of Born-Oppeheimer approximation is presented in \cite{combes-seiler}. It is not at all clear that the eigenvalues of the light degrees of freedom, that one computes, under the influence of potentials when the heavy centers are clamped,  actually define well-behaved nonintersecting surfaces  when one considers the heavy degrees of freedoms as parameters. This difficult problem is solved by Hunziker in \cite{hunziker}, where even for Coulomb type potentials energy eigenfunctions are shown to be essentially analytic functions of the heavy coordinates. These problems further investigated in a series of papers by Hagedorn \cite{hagedorn4,hagedorn5,hagedorn6}. In the usual potentials, higher order corrections to the Born-Oppenheimer approximation  are rigorously investigated by Hagedorn in a series of papers \cite{hagedorn1,hagedorn2}.  Further investigations along similar lines are presented in \cite{seiler-klein}.  They typically correspond to higher order corrections to the effective potentials generated by the light degrees of freedom.  A different attempt to include higher order corrections to Born-Oppenheimer approximation is given by Weingert and Littlejohn \cite{weigert} as an example of their diagonalization technique in the deformation quantization approach. They discover derivative terms in the corrections, in our problem especially being a highly singular system, we get kinetic energy corrections even at the leading order. Therefore, the common wisdom does not seem to work in this system. In our work, our results also indicate the importance of such kinetic energy corrections. The reader can find a large collection of references and various  mathematically precise statements on Born-Oppenheimer approximation in a recent review by Jecko \cite{Jecko-5}.

In a similar spirit to \cite{seiler-2}, a slightly simpler model for its pedagocial value was proposed by G. Gangopadhyay and B. Dutta-Roy in \cite{dutta-roy} where the authors consider a light particle coupled to a heavy particle via a delta function potential and the whole system is confined to a box in one dimension. This system has a distributional potential yet it is not truly singular, since an analytical treatment is still possible one can test the Born-Oppenheimer approximation.  The problem we propose to consider  is such a toy model in a highly nontrivial sitution, which requires a coupling constant renormalization. Recently, we consider a simpler version of this problem in one dimension, in which there is no renormalization, however we could analyse this case in much more detail \cite{haci-turgut}.
Born-Oppenheimer Approximation is  a well-known approach for treating coupling of fast (light) and slow (heavy) degrees of freedom in typical quantum mechanical systems. However the existing literature is tailored to deal with interaction potentials which are sufficiently regular. It is not so obvious that these methods would suffice to deal with somewhat more singular cases, especially the problems which require renormalization. In view of the fact that most quantum field theories require renormalization this is an especially important problem which has direct physical implications. Insight to be gained in some simple versions may shed light on the methods to be developed for more realistic systems. The model we have in mind looks very simple, we consider two heavy particles in two dimensions interacting via a contact term with a third light particle. To make the calculations tractable we use a nonrelativistic approach, nevertheless a more interesting verison would be to treat the light particle as a relativistic one. The relativistic model may be relevant for understanding the following problem, just as in the case of nucleons interacting with each other via exchange of mesons, interaction of  two heavy quarks may be energetically more favorable by  exchange of glueballs, which are to be modeled as light mass relativistic particles. 
In any case, we see that the basic idea of separating heavy (slow) and light (fast) modes does indeed work, yet in a rather nontrivial manner. One does not find a simple expansion scheme, where corrections manifest themselves easily. It has to be worked out carefully, and even the leading term requires delicate set of computations. We see that the leading term is actually {\it large}, and dominates the energy of the system, this is an effect due to the nonlinear nature of renormalization. One may compute some of the lower order terms, some contain logarithm of the large parameter $M/m$, there are also lower order terms one can in principle compute. A many body approach could be more natural to develop a systematic expansion.

\section{Two heavy particles interacting with a light one}
 
We write down the Schr\"odinger equation for this simple model;
\begin{equation}
\Big[-\frac{\hbar^2}{2M}\sum_{i}{\nabla_i}^2-\frac{\hbar^2}{2m}\nabla^2-\lambda\delta(x-x_1)-\lambda\delta(x-x_2)]\Big]\Psi(x;x_1, x_2)=E\Psi(x;x_1,x_2).
\end{equation}
Here $x_1,x_2$ refer to the heavy particles coordinates and $x$ refers to the light one. The choices of the masses also reflect this difference. 
Let us pretend that the Born-Oppenheimer approximation can be applied to this system, thus we use  a decomposition of the wave function into fast and
slow degrees of freedom:
\begin{equation}
\Psi(x;x_1, x_2)=\phi(x|x_1, x_2)\psi(x_1,x_2).
\end{equation}
We assume that this decomposition respects the translational invariance of the system, moreover we assume for the time being that the delta-functions are actually regularized since we know tha there is a divergence hidden in this problem.
We substitute the proposed solution into  the Schr\"odinger equation, 
\begin{eqnarray}
&\ & \Big[-\frac{\hbar^2}{2M}\sum_{i}{\nabla_i}^2\psi(x_1,x_2)\Big]\phi(x|x_1,x_2)+ \Big[-\frac{\hbar^2}{2M}\sum_{i}{\nabla_i}^2\phi(x|x_1,x_2)\Big]\psi(x_1,x_2) \nonumber\\
&\ & -\frac{\hbar^2}{2M}\sum_{i} {\partial \phi\over \partial x_i}{\partial \psi\over \partial x_i}
  +\Big[\big(-\frac{\hbar^2}{2m}\nabla_x^2-\lambda\delta(x-x_1)-\lambda\delta(x-x_2)\big)\phi(x |x_1,x_2)\Big]\psi(x_1, x_2)\nonumber\\
&\  & \quad \quad \quad \quad\ \ \ \ \ \ \ \ \ \ \ \ \ \ \ \ \ \ \  \ \ \ \ \ \ \ \ \ \ \ \quad =E\phi(x|x_1,x_2)\psi(x_1,x_2).
\end{eqnarray}
Ordinarily we  would assume  that we could find the solution to the equation below,
\begin{equation}
-\frac{\hbar^2}{2m}\nabla_x^2 \phi(x|x_1, x_2)-\lambda[\delta(x-x_1)+\delta(x-x_2)]\phi(x|x_1, x_2) =E(x_1,x_2)\phi(x|x_1, x_2)
.\end{equation}
That woul mean that  the heavy particles would act like fixed centers and the light particle would move in this background.
Then we would find,
\begin{eqnarray}
&\ & \ \ \Big[-\frac{\hbar^2}{2M}\sum_{i}{\nabla_i}^2\psi(x_1,x_2)+E(x_1,x_2)\psi(x_1,x_2)\Big]\phi(x|x_1,x_2)-\frac{\hbar^2}{2M}\sum_{i} \nabla_i \phi\cdot \nabla_i \psi\nonumber\\
&\ &  +\Big[-\frac{\hbar^2}{2M}\sum_{i}{\nabla_i}^2\phi(x|x_1,x_2)\Big]\psi(x_1,x_2)  =E\phi(x|x_1,x_2)\psi(x_1,x_2)
,\end{eqnarray}
 and if we could neglect the last two terms on the lefthand side, we would end up with the Born-Oppenheimer result,
\begin{equation}
\Big[-\frac{\hbar^2}{2M}\sum_{i}{\nabla_i}^2+E(x_1,x_2)\Big]\psi(x_1,x_2)=E\psi(x_1,x_2)
.\end{equation} 
Nevertheless we will see that this would be wrong, and we would end up with a divergent result. The expression
\begin{equation}
\Big[-\frac{\hbar^2}{2M}\sum_{i}{\nabla_i}^2\phi(x|x_1,x_2)\Big]
,\end{equation}
contains a term that we can convert into
\begin{equation}
\Big[-\frac{\hbar^2}{2M}{\nabla_x}^2\phi(x|x_1,x_2)\Big]
\end{equation}
hence should be incorporated into the term (4), resulting into 
\begin{equation}
-\Big(\frac{\hbar^2}{2m}+\frac{\hbar^2}{2M}\Big)\nabla_x^2 \phi(x|x_1, x_2)-\lambda[\delta(x-x_1)+\delta(x-x_2)]\phi(x|x_1, x_2) =E(x_1,x_2)\phi(x|x_1, x_2)
,\end{equation}
to be renormalized all together in this two dimensional case. Note that because of the divergence, the usual rule of ignoring $m/M$ terms at this order does not work here. Moreover, the effective potential generated from the derivative terms contain a ${1\over z^2}$ term, where $z$ is the relative coordinate for the two heavy particles. This term cannot be added as a perturbation since it changes the character of the wave function at the origin, hence should be used in the leading order Born-Oppenheimer approximation. 
 
To set up the formalism, we will introduce an ansatz for the solution of the light degrees of freedom, assuming for the time being that the delta functions are properly regularized--one possibility is to use the heat kernel itself, this will preserve the translational invariance of the whole  system, for the time being we will proceed formally; 
\begin{equation}
\phi(x|x_1,x_2)=A(x_1,x_2)\Big(\eta_+(x|x_1,x_2)+\eta_-(x|x_1,x_2)\Big)
.\end{equation}
Later, we use as coordinates, the center of mass of the heavy particles and the relative position, 
$$
X=\frac{x_1+x_2}{2}\quad {\rm and} \quad z=x_1-x_2,
$$
we will see that these are the natural coordinates for our system.
Let us search for the solution in the following form,
$$
\phi(x|x_1,x_2)={N}\left[\int_0^\infty\frac{ dt}{\hbar} K_{ t}(x,x_1)e^{-\frac{\nu^2}{\hbar}t}+\int_0^\infty \frac{dt}{\hbar} K_{ t}(x,x_2)e^{-\frac{\nu^2}{\hbar}t}\right],
$$
where 
\begin{equation}
-{\hbar^2\over 2m_*}\nabla_x^2 K_t(x,y)+\hbar {\partial K_t(x,y)\over \partial t}=0
,\end{equation}
 is the usual heat equation in two-dimensions corresponding to a mass,
\begin{equation}
{1\over m_*}= {1\over m}+{1\over M}
,\end{equation}
the solution of which is the well-known Gaussian when we demand that it goes to  a delta-function as time goes to zero.  Here,  with the mentioned divergence in mind, we choose a corrected mass $m_*$ for the heat kernel. Consequently to get a solution (after the cut-off being removed), we need to satisfy the equation, 
$$
\frac1{\lambda}-\frac1{\hbar}\int_0^\infty dt K_{ t}(x_1,x_1)e^{-\frac{\nu^2}{\hbar}t}=\frac1{\hbar}\int_0^\infty dt K_{ t}(x_1,x_2)e^{-\frac{\nu^2}{\hbar}t}.
$$
As a result  to cure the divergence we need to choose the coupling constant as,
$$
\frac1{\lambda}=\frac1{\hbar}\int_0^\infty dt K_t(x_1,x_1)e^{-\frac{\epsilon^2}{\hbar}t},
$$
in which an arbitrary bound state energy $\epsilon^2$ appears for the system. This is the binding energy we need to decide, when a single heavy center and a light particle interact via a contact term,  just as one needs to determine the proper coupling constant for a given physical system by some measurement we assume that the binding energy is the measured quantity. The energy of two heavy particles case  is to be determined from this input.  Here the dimensionless coupling constant is to be traded over with this binding energy, for this particular choice there is no need to introduce a finite part to the coupling constant. Thus we find a well-defined expression,
$$
\frac{2m_*}{4\pi\hbar^2}\int_0^\infty \frac{dt}{t}(e^{-\frac{\epsilon^2}{\hbar}t}-e^{-\frac{\nu^2}{\hbar}t})=\frac{2m_*}{4\pi\hbar^2}\int_0^\infty \frac{dt}{t}e^{-\frac{\nu^2}{\hbar}t-\frac{2m_*z^2}{4\hbar t}}.
$$
Consequently we find the equation to be satisfied for the binding energy of the total system, in this approximation,
$$
\ln\left(\frac{\nu^2}{\epsilon^2}\right)=2K_0\left(\frac{\sqrt{2m_*}}{\hbar}\nu|z|\right),
$$
where $K_0(.)$ is refers to the well-known modified Bessel function.
Let us keep this expression in mind and calculate the resulting normalized wave functions, this requires evaluating,
$$
1=\frac{N^2}{\hbar^2}\left[\int_0^\infty dt_1dt_2[K_{t_1+t_2}(x_1,x_1)+K_{t_1+t_2}(x_2,x_2)]e^{-\frac{\nu^2}{\hbar}(t_1+t_2)}+2 \int_0^\infty dt_1dt_2K_{t_1+t_2}(x_1,x_2)e^{-\frac{\nu^2}{\hbar}(t_1+t_2)}\right]
,$$
where we use the reproducing property of the Gaussian expression for the heat kernels. The integrals can be done easily by transforming to the variables,
$$
t=t_1+t_2,s=t_1-t_2
$$
$$
1=\frac{N^2}{\hbar^2}\left[\int_0^\infty dt te^{-\frac{\nu^2}{\hbar}t}\left[K_t(x_1,x_1)+K_t(x_2,x_2)\right]+2\int_0^\infty dt tK_t(x_1,x_2)e^{-\frac{\nu^2}{\hbar}t}\right]
,$$
or equivalently,
$$
1=2\frac{N^2}{\hbar^2}\left(\frac{2m_*}{4\pi\hbar}\right)\left[\int_0^\infty dte^{-\frac{\nu^2}{\hbar}t}+\int_0^\infty dte^{-\frac{\nu^2}{\hbar}t-\frac{2m_*z^2}{4\hbar t}}\right]
.$$
This gives us,
$$
N=\frac1{\sqrt{2}}\sqrt{\hbar}\nu\sqrt{\frac{4\pi\hbar}{2m_*}}\frac1{\sqrt{\left[1+\frac{\sqrt{2m_*}}{\hbar}\nu|z|K_1\left(\frac{\sqrt{2m_*}}{\hbar}\nu|z|\right)\right]}}
,$$
where $K_1(.)$ refers to the modified Bessel function. Consequently, we have the normalized wave function,
$$
\phi(x|x_1,x_2)=\frac1{\sqrt{2\pi}}\frac{\frac{\sqrt{2m_*}}{\hbar}\nu}{\sqrt{\left[1+\frac{\sqrt{2m_*}}{\hbar}\nu|z|K_1\left(\frac{\sqrt{2m_*}}{\hbar}\nu|z|\right)\right]}}\left[K_0\left(\frac{\sqrt{2m_*}}{\hbar}\nu|x-x_1|\right)+K_0\left(\frac{\sqrt{2m_*}}{\hbar}\nu|x-x_2|\right)\right]
$$
where $K_0(.)$ again is the zeroth order modified Bessel function. Using the more natural variables, 
$X=\frac{x_1+x_2}{2}$ and $z=x_1-x_2$, we can rewrite,
$$
\phi(x|X,z)=\frac{{1\over \sqrt{2\pi}}\frac{\sqrt{2m_*}}{\hbar}\nu}{\sqrt{\left[1+\frac{\sqrt{2m_*}}{\hbar}\nu|z|K_1\left(\frac{\sqrt{2m_*}}{\hbar}\nu|z|\right)\right]}}\left[K_0\left(\frac{\sqrt{2m_*}}{\hbar}\nu|x-X+\frac{z}{2}|\right)+K_0\left(\frac{\sqrt{2m_*}}{\hbar}\nu|x-X-\frac{z}{2}|\right)\right]
$$
$$
       =A(z)\Big[\eta_+(\nu(z),x-X+z/2)+\eta_-(\nu(z),x-X-z/2)\Big].
$$
Here we use,
\begin{eqnarray}
A(z)&=&\frac1{\sqrt{2\pi}}\frac{\sqrt{2m_*}}{\hbar}\nu\frac1{\sqrt{1+\frac{\sqrt{2m_*}}{\hbar}\nu|z|K_1\left(\frac{\sqrt{2m_*}}{\hbar}\nu|z|\right)}}\nonumber\\
\eta_\pm(\nu, x-X\pm z/2)&=&K_0\left(\frac{\sqrt{2m_*}}{\hbar}\nu\left|x-X\pm\frac{z}{2}\right|\right).\nonumber
\end{eqnarray} 
Let us emphasize again that here $\nu=\nu(|z|)$, so it also depends on the distance $|z|$ between the two heavy centers and this will be crucial in our computations. As we will see, heavy particle limit implies that 
$|z|$ can be considered as small relative to the light particle length scale to be made precise below.

We now take  a step back and write equation (3) in the new coordinates with the proposed wave functions in mind,
\begin{eqnarray}
&\ & \Big[-\frac{\hbar^2}{4M}{\nabla_X^2}-\frac{\hbar^2}{M}{\nabla_z^2}\Big]\psi(X,z)\phi(x|X,z)
  +\Big[\big(-\frac{\hbar^2}{2m}\nabla_x^2-\lambda\delta(x-x_1)-\lambda\delta(x-x_2)\big)\phi(x |X,z)\Big]\psi(X,z)\nonumber\\
&\  & \quad \quad \quad \quad\ \ \ \ \ \ \ \ \ \ \ \ \ \ \ \ \ \ \  \ \ \ \ \ \ \ \ \ \ \ \quad =E\phi(x|X,z)\psi(X,z).
\end{eqnarray}
We can write this as,
\begin{eqnarray}
&\ & \Bigg(\Big[-\frac{\hbar^2}{4M}{\nabla_X^2}-\frac{\hbar^2}{M}{\nabla_z^2}\Big]\psi(X,z)\Bigg)\phi(x|X,z) +\Bigg(\Big[-\frac{\hbar^2}{4M}{\nabla_X^2}-\frac{\hbar^2}{M}{\nabla_z^2}\Big]\phi(x|X,z)\Bigg)\psi(X,z)\nonumber\\
&\ &\quad\quad \quad  -\frac{\hbar^2}{4M}{\nabla_X\psi(X,z)}\cdot\nabla_X \phi(x|X,z)-\frac{\hbar^2}{M}{\nabla_z\psi(X,z)}\cdot\nabla_z \phi(x|X,z)\nonumber\\
&\ &  \quad\quad \quad +\Big[\big(-\frac{\hbar^2}{2m}\nabla_x^2-\lambda\delta(x-x_1)-\lambda\delta(x-x_2)\big)\phi(x |X,z)\Big]\psi(X,z) =E\phi(x|X,z)\psi(X,z).\nonumber
\end{eqnarray}
Let us consider the second term and using the decomposition of the wave function we see that the first derivative becomes,
\begin{eqnarray}
&\ & (-\frac{\hbar^2}{4M}{\nabla_X^2})\phi(x|X,z)=A(z)(-\frac{\hbar^2}{4M}{\nabla_X^2})[\eta_+(\nu(z),x-X+z/2)+\eta_-(\nu(z),x-X-z/2)]\nonumber\\
&\ & \quad \quad \quad =A(z)(-\frac{\hbar^2}{4M}{\nabla_x^2})[\eta_+(\nu(z),x-X+z/2)+\eta_-(\nu(z),x-X-z/2)]\nonumber\\
&\ & \quad \quad =(-\frac{\hbar^2}{4M}{\nabla_x^2})\phi(x|X,z).
\end{eqnarray}
The other derivative requires more care, let us note that the functions $\eta_\pm$ depend on $z$ in two ways, one is through the difference $x-X\pm z/2$ the  other is through the term $\nu(z)$.
It is the first dependence that we should pay more attention, let us  divide the $z$ derivatives acting on these functions as follows,
$$
\nabla_z=\nabla_z|_\nu+(\nabla_z \nu) {\partial\over \partial \nu}.
$$
Therefore, we have 
$$
\nabla_z^2=\nabla_z^2|_\nu+(\nabla_z^2 \nu){\partial\over \partial \nu}+(\nabla_z \nu)^2 {\partial^2\over \partial \nu^2}+2(\nabla_z \nu) {\partial\over \partial \nu}\nabla_z|_\nu.
$$ 
Again, the important term here is the first one,
acting on the $\eta_\pm$ part of the wave function,
\begin{equation}
\nabla_z^2|_\nu(\eta_++\eta_-)={1\over 4} (\nabla_x^2\eta_++\nabla^2_x\eta_-)
,\end{equation}
because, $\nabla_z \eta_\pm=\pm {1\over 2} \nabla_x \eta_\pm$.
Hence, we have this part, named as the singular part,  to be separated from the full system as claimed, 
\begin{equation}
\Big[-\frac{\hbar^2}{ M}\nabla_z^2\phi(x|X,z)\Big]_{\rm sing}=-\frac{\hbar^2}{ 4M}\nabla_x^2\phi(x|X,z).
\end{equation}
As a result, we have
\begin{eqnarray}
&\ & \Bigg(\Big[-\frac{\hbar^2}{4M}\nabla_X^2-\frac{\hbar^2}{M}{\nabla_z^2}\Big]\psi(X,z)\Bigg)\phi(x|X,z) +\Bigg(\Big[-\frac{\hbar^2}{M}{\nabla_z^2}\Big]_{\rm reg}\phi(x|X,z)\Bigg)\psi(X,z)\nonumber\\
&\ &\quad\quad \quad  -\frac{\hbar^2}{4M}{\nabla_X\psi(X,z)}\cdot\nabla_X \phi(x|X,z)-\frac{\hbar^2}{M}{\nabla_z\psi(X,z)}\cdot\nabla_z \phi(x|X,z)\nonumber\\
&\ &  \quad\quad \quad +\Bigg(\Big[-\frac{\hbar^2}{2M}\nabla_x^2-\frac{\hbar^2}{2m}\nabla_x^2-\lambda\delta(x-x_1)-\lambda\delta(x-x_2)\Big]\phi(x |X,z)\Bigg)\psi(X,z) =E\phi(x|X,z)\psi(X,z).\nonumber
\end{eqnarray}
Here, we named the remaining piece as the regular part.
To simplify the computations we may assume $\psi(X,z)=e^{iQX}\Psi(z)$, that would essentially remove all the center of mass terms of the heavy particles from this expression, and we take the expectation value with the light degrees of freedom:
\begin{eqnarray}
&\ & \int dx \phi(x|X,z)\Bigg[\Big(-\frac{\hbar^2}{M}{\nabla_z^2}\Psi(z)\Big)\phi(x|X,z) +\Bigg(\Big[-\frac{\hbar^2}{M}{\nabla_z^2}\Big]_{\rm reg}\phi(x|X,z)\Bigg)\Psi(z)\nonumber\\
&\ &\quad\quad \quad  - \frac{\hbar^2}{4M}{\Psi(z)}Q\cdot\nabla_X \phi(x|X,z)-\frac{\hbar^2}{M}{\nabla_z\Psi(z)}\cdot\nabla_z \phi(x|X,z)\Bigg]\nonumber\\
&\ &  \quad\quad \quad +\underbrace{\int dx \Bigg(\phi(x|X,z)\Big[-\frac{\hbar^2}{2M}\nabla_x^2-\frac{\hbar^2}{2m}\nabla_x^2-\lambda\delta(x-x_1)-\lambda\delta(x-x_2)\Big]\phi(x |X,z)\Bigg)}_{-\nu^2(|z|) \quad {\rm after \ renormalization}}\Psi(z)\nonumber\\
&\ & \quad\quad \quad \quad \quad\quad \quad \quad \quad  =\Big[E-{\hbar^2Q^2\over 4M}\Big]\Psi(z)=\delta E \Psi(z).\nonumber
\end{eqnarray}
Thus we should work out all the individual terms here, we will see that the effective potential does not only come from the expansion of the binding energy $-\nu^2(|z|)$, the cross terms will also matter, especially the repulsive ${1\over |z|^2}$ generated by these terms will make the wave function to vanish  mildly at the origin. As a result,
\begin{eqnarray}
&\ &\!\!\!\!\!\! -\frac{\hbar^2}{M}{\nabla_z^2}\Psi(z)-\nu^2(|z|)\Psi(z)-\frac{\hbar^2}{M}\int dx \phi(x-X|z)\nabla_z \phi(x-X|z)\cdot\nabla_z \Psi(z) \nonumber\\
&\ &\quad  +\Bigg(\int dx \phi(x-X|z)\Big[-\frac{\hbar^2}{M}{\nabla_z^2}\Big]_{\rm reg}\phi(x-X|z)- \frac{\hbar^2}{4M}\int dx \phi(x-X|z)Q\cdot\nabla_X \phi(x-X|z)\Bigg) \Psi(z)\nonumber\\
&\ & \quad\quad \quad \quad \quad\quad \quad \quad \quad =\delta E \Psi(z),\nonumber
\end{eqnarray}
note that we deliberately wrote $\phi(x|X,z)=\phi(x-X|z)$ to emphasize this translational invariance of $x$, as a result $\nabla_X \phi(x-X|z)=-\nabla_x\phi(x-X|z)$, furthermore,  we can redefine $x\to x-X$, hence the apparent dependence on $X$ dissappears, we end up with,
\begin{eqnarray}
&\ &\!\!\!\!\!\!\!\!\! -\frac{\hbar^2}{M}{\nabla_z^2}\Psi(z)-\nu^2(|z|)\Psi(z)-\frac{\hbar^2}{M}\int d^2x \phi(x|z)\nabla_z \phi(x|z)\cdot\nabla_z \Psi(z) \nonumber\\
&\ &\quad  +\Bigg(\int d^2x \phi(x|z)\Big[-\frac{\hbar^2}{M}{\nabla_z^2}\Big]_{\rm reg}\phi(x|z)+ \frac{\hbar^2}{4M}Q\cdot \underbrace{\int d^2x \phi(x|z)\nabla_x \phi(x|z)}_{{\rm being\ an\  exact\  differential}=0}\Bigg) \Psi(z)\nonumber\\
&\ & \qquad \qquad \qquad \qquad \qquad \qquad \qquad \quad \qquad =\delta E \Psi(z).\label{averaged}
\end{eqnarray}
In the subsequent pages we will complete this delicate calculation and obtain the effective potential, let us reiterate that the seemingly negligible character of these terms are misleading, the resulting potential being rather singular and repulsive,  this influences the behaviour of the wave function significantly around the origin. As we will see, $\delta E$ {\it is not small}, it is the length scale associated to the heavy particle pair which is small compared to the light particle's spread of the wave function.
The reader who is interested in the resulting potential can skip these calculations; they are essentially of a technical nature.
 
Let us start with the regular expression;
\begin{eqnarray}
\int dx \phi(x|z)\Big[-\frac{\hbar^2}{M}{\nabla_z^2}\Big]_{\rm reg}\phi(x|z),
\end{eqnarray}
we recall that,
\begin{equation}
\phi(x|z)=A(z)\Big(\eta_+(x+z/2)+\eta_-(x+z/2)\Big).
\end{equation}
Let us explicitly write this expectation value,
\begin{eqnarray}
&\ &-\frac{\hbar^2}{M}\int dx A(z)(\eta_++\eta_-)\Big[\nabla_z^2\Big(A(z)(\eta_++\eta_-)\Big)\Big]_{\rm reg}\nonumber\\
&=&-\frac{\hbar^2}{M}\int d^2 xA(z)(\eta_++\eta_-)\Bigg[\underbrace{A(z)\Big[\nabla_z^2(\eta_++\eta_-)\Big]_{\rm reg}}_{(1)}+\underbrace{(\eta_++\eta_-)\nabla_z^2A(z)}_{(2)}+\underbrace{2{\nabla_z}A(z)\cdot{\nabla_z}(\eta_++\eta_-)}_{(3)}\Bigg]\nonumber
\end{eqnarray}
so after the integration  the numbered terms produce the following results:
\begin{eqnarray}
(1) &=&-\frac{\hbar^2}{M}A^2\int d^2x(\eta_++\eta_-)\Big[\nabla_z^2(\eta_++\eta_-)\Big]_{\rm reg}\nonumber\\
(2)&=&-\frac{\hbar^2}{M}\left(\frac1{A}\nabla_z^2A\right)\nonumber\\
(3) &=&-\frac{\hbar^2}{M/2}A\int d^2x (\eta_++\eta_-){\nabla_z}A\cdot{\nabla_z}(\eta_++\eta_-)
\end{eqnarray}
Let us recall that acting on $\eta_\pm$ the regular part of the $z$ derivatives are given by 
$$
[\nabla_z^2]_{\rm reg}\eta_\pm=[(\nabla_z^2 \nu){\partial\over \partial \nu}+(\nabla_z \nu)^2 {\partial^2\over \partial \nu^2}+2(\nabla_z \nu) \cdot {\partial\over \partial \nu}\nabla_z|_\nu]\eta_\pm,
$$ 
here in {\it cylindrical coordinates} $z,\theta$, $\nabla_z^2={\partial^2\over \partial z^2}+{1\over z}{\partial \over \partial z}$, note that \textcolor{red}{\it not to complicate the notation we use the same letter} $z$ \textcolor{red}{\it 
for the radial coordinate}. Hopefully this will not lead to any confusion, since pure $z$ derivatives only appear as $\nabla_z$, and usually converted into radial derivatives.  
Note that this expression can be equivalently written,
$$
[\nabla_z^2]_{\rm reg}\eta_\pm=\left[(\nabla_z^2 \nu){\partial\over \partial \nu}+({\partial \nu \over \partial z})^2 {\partial^2\over \partial \nu^2}+2{\partial\nu\over\partial z}  {\partial\over \partial \nu}\hat z\cdot \nabla_z|_\nu\right]\eta_\pm,
$$ 
this is the form that we will be using.
For some of the calculations to follow, we  remind the equation satisfied by $\nu$ below,
\begin{equation}
\ln\left(\frac{\nu^2}{\epsilon^2}\right)=2K_0\left(\frac{\sqrt{2m}}{\hbar}\nu z\right)\nonumber
\end{equation}

As a crucial step in Born-Oppenheimer approximation,
 we {\it assume} that $\frac{\sqrt{2m_*}}{\hbar}\epsilon z<<1$, the validity of which is to be justified later.
We note that when  $z$  is small compared to the distance scale defined by the wave function of the light particle, $\sqrt{2m_*}\epsilon/\hbar$, we may expand the equation giving the bound state energy $\nu$ as a function of $z$, using the well-known short distance behaviour of $K_0(x)\approx -\ln(xe^\gamma/2)+...$, 
$$
\ln\left(\frac{\nu^2}{\epsilon^2}\right)=-2\ln\left(\frac{\sqrt{2m_*}}{2\hbar}\nu ze^\gamma\right)-2\ln\left(\frac{\sqrt{2m_*}}{2\hbar}\nu ze^{\gamma-1}\right)\frac{2m_*}{4\hbar^2}\nu^2z^2+...,
$$
Then we collect the similar terms together,
\begin{eqnarray}
&\ &2\ln\left(\frac{\nu^2}{\epsilon^2}\right)\left[1+\frac1{8}\frac{2m_*}{\hbar^2}\nu^2z^2+...\right]\nonumber\\
&\ & =-2\ln\left(\frac{\sqrt{2m_*}}{2\hbar}\epsilon ze^\gamma\right)-2\ln\left(\frac{\sqrt{2m_*}}{2\hbar}\epsilon ze^{\gamma-1}\right)\frac{2m_*}{4\hbar^2}\nu^2 z^2+...\nonumber,
\end{eqnarray}
note that in the second logarithm  term the factor $e^{\gamma-1}$ can be replaced with $e^\gamma$ since this constant factor being multiplied with $z^2$ is of lower order, as long as we keep up to the second order expansion, so to simplify the expression we change this term to $e^\gamma$. Hence we can rewrite all of it as,
$$
\ln\left(\frac{\nu^2}{\epsilon^2}\right)=-\left[\frac{1+\frac{2m_*}{4\hbar^2}\nu^2z^2+...}{1+\frac{2m_*}{8\hbar^2}\nu^2z^2+...}\right]\ln\left(\frac{\sqrt{2m_*}}{2\hbar}\epsilon ze^\gamma\right).
$$
By expanding again the denominator,
$$
\ln\left(\frac{\nu^2}{\epsilon^2}\right)=-\left[1+\frac{2m_*}{8\hbar^2}\nu^2z^2...\right]\ln\left(\frac{\sqrt{2m_*}}{2\hbar}\epsilon ze^\gamma\right).
$$
Let us reorganize this expression, to this purpose we define
\begin{equation}
\xi=\left(\frac{\nu^2}{\epsilon^2}\right)\frac{\sqrt{2m_*}}{\hbar}\epsilon z,
\end{equation}
then,
$$
\ln\left(\xi \frac{e^\gamma}{2}\right)=-\frac1{4}\xi e^{-\gamma}x\ln x+...,
$$
here we introduced $x=\frac{\sqrt{2m_*}}{2\hbar}\epsilon z e^\gamma$. 
This equation can be solved order by order by iteration, since  $x\ln(x)<<1$, we see that  $\xi\approx{2 e^{-\gamma}}$, to a first approximation, which defines our zeroth order expression. We will be content with the first order corrected result only, 
$$
\nu^2\approx\frac{\hbar}{\sqrt{2m_*}}\frac{2\epsilon}{ze^\gamma}e^{-\frac{e^{-2\gamma}}{2}x\ln x}=\frac{\hbar}{\sqrt{2m_*}}\frac{2\epsilon}{z e^{\gamma}}\left(1-\frac{e^{-\gamma}}{4}\frac{\sqrt{2m_*}}{\hbar}\epsilon z \ln\left[\frac{\sqrt{2m_*}}{\hbar}\epsilon z\right]+...\right),
$$
where we dropped the precise constant in the logarithmic term. Notice that  we have the emergence of  the characteristic length scale $\zeta_0={\hbar\over\sqrt{2m_*} \epsilon}$, which describes the spread of the light particle-heavy particle bound state wave function in our calculations. 
  
Knowing the exact relation for $\nu$, one  can immediately find the resulting derivative of $\nu$ with respect to the variable $z$:
\begin{equation}
\frac{\partial\nu}{\partial z}=-\frac{\frac{\sqrt{2m_*}}{\hbar}\nu^2 K_1\left(\frac{\sqrt{2m_*}}{\hbar}\nu z\right)}{\left[1+\frac{\sqrt{2m_*}}{\hbar}\nu z K_1\left(\frac{\sqrt{2m_*}}{\hbar}\nu z \right)\right]}.
\end{equation}
We remind for the convenience of the reader,  small argument expansion of Bessel function $K_1(x)$,
\begin{equation}
K_1(x)\approx {1\over 2} {1\over (x/2) }+{1\over 2} x \ln({x\over 2})+...\quad {\rm as} \ x\to 0^+
,\end{equation}
the following term is of order $x$ which we neglect, however this leads to an ambiguity in the logarithmic part, any multiplicative constant can be admitted in the argument since it only affects the result at the neglected  next order.
At small distance, using the above approximation,  we find,
\begin{equation}
\frac{\partial\nu}{\partial z}\approx -\frac1{2}\frac{\nu}{z}-{1\over 8}\Big(\frac{\sqrt{2m_*}}{\hbar}\Big)^2\nu^3z\ln \left[\frac{\sqrt{2m_*}}{2\hbar}\nu z\right] \label{nufirstder}
,\end{equation}
which should be further simplified by using the expansion of $\nu(z)$ to first order. For the time being we only need the zeroth order expressions, so we will keep it as it is given.
We also need the expression for the second derivative, which we can find exactly and present in an Appendix, we write here the leading order expansion,  using the {\it radial variable} $z$,  as follows
\begin{equation}
{\partial^2 \nu\over \partial z^2}\approx {3\over 4} {\nu\over z^2}+... \label{nusecondder}
,\end{equation}
note that here we should expand $\nu$ to first order again to find a consistent expansion. We remark that {\it for simplicity we kept} $m_*$ {\it as it is}, it can be replaced with $m$ at various places, but we may do this at the end.

Let us go back and work out each term separately, as we numbered them,
the first piece becomes the following;
\begin{eqnarray}
(1) &=& -\frac{\hbar^2}{M}A^2\int d^2x(\eta_++\eta_-)\Big(2{\nabla_z\nu}\cdot \nabla_z|_\nu\partial_\nu+\Big[{1\over z} {\partial \nu\over \partial z} +\frac{\partial^2\nu}{\partial z^2}\Big]\partial_\nu+\left(\frac{\partial\nu}{\partial z}\right)^2\partial_\nu^2\Big)(\eta_++\eta_-)\nonumber\\
& =& -\frac{\hbar^2}{M}A^2\int d^2x \Bigg[\nabla_z \nu \cdot \eta_+\nabla_x \partial_\nu\eta_+ -\nabla_z \nu \cdot \eta_-\nabla_x\partial_\nu\eta_-+\underbrace{\nabla_z\nu \cdot \eta_-\nabla_x\partial_\nu\eta_+ -\nabla_z \nu \cdot \eta_+\nabla_x\partial_\nu\eta_-}_{(1_b)}\nonumber\\
&+& \underbrace{(\eta_++\eta_-)\Big[{1\over z} {\partial \nu\over \partial z} +\frac{\partial^2\nu}{\partial z^2}\Big]\partial_\nu(\eta_++\eta_-)}_{(1_c)}+\underbrace{(\eta_++\eta_-)\left(\frac{\partial\nu}{\partial z}\right)^2\partial_\nu^2(\eta_++\eta_-)}_{(1_d)}\Bigg],
\end{eqnarray}
note that we used $\nabla_{z}|_\nu \eta_\pm=\pm{1\over 2}\nabla_x \eta_\pm$ in the first few terms.
Shifting the integration variable by $\pm z/2$, one can check that,
\begin{equation}
\int d^2x\left[\nabla_z \nu \cdot \eta_+\nabla_x \partial_\nu\eta_+ -\nabla_z \nu \cdot \eta_-\nabla_x\partial_\nu\eta_-\right]=0.
\end{equation}
Let us consider the other cross term, by an explicit computation of the derivative terms we find out that,
\begin{eqnarray}
&(1_b) &=-\frac{\hbar^2}{M}A^2\int d^2x\left[\eta_-\frac{\partial\nu}{\partial z}\hat z \cdot \nabla_x\partial_\nu\eta_+ -\eta_+\frac{\partial\nu}{\partial z}\hat z \cdot \nabla_x\partial_\nu\eta_-\right]\nonumber\\
&\ &=-\frac{\hbar^2}{M}A^2\frac{2m_*}{\hbar^2}\nu z \frac{\partial\nu}{\partial z}\int d^2x\eta_+\eta_-\nonumber\\
&\  &=-\frac{\hbar^2}{M}A^2\frac{2m_*}{\hbar^2}\nu z\frac{\partial\nu}{\partial z}\frac{\pi\hbar}{2m_*}\frac{\sqrt{2m_*}}{\nu}zK_1\left(\frac{\sqrt{2m_*}}{\hbar}\nu z\right).
\end{eqnarray}
Up to this point we have an exact calculation. Let us now expand each term to leading order under the small $z/\zeta_0$ assumption as before,
\begin{eqnarray}
&(1_b) &\approx-\frac{2m_*}{M}A^2\frac{\partial\nu}{\partial z}\frac{z}{\nu}\frac{\pi\hbar^2}{2m_*}\approx \frac1{2}\frac{m_*}{M}\nu^2\approx\frac{1}{2}\frac{m_*}{M}\frac{\hbar}{\sqrt{2m_*}}\frac{\epsilon}{z}=\frac{1}{2}\frac{m_*}{M}\epsilon^2\frac{\zeta_0}{z},
\end{eqnarray}
which shows that it is a small perturbation to the leading terms.
We now calculate the next term,
\begin{eqnarray}
(1_c)&=&-\frac{\hbar^2}{M}A^2\Big[{1\over z} {\partial \nu\over \partial z} +\frac{\partial^2\nu}{\partial z^2}\Big]\int d^2x(\eta_++\eta_-)\partial_\nu(\eta_++\eta_-)\nonumber\\
&=&-\frac{\hbar^2}{2M}A^2\Big[{1\over z} {\partial \nu\over \partial z} +\frac{\partial^2\nu}{\partial z^2}\Big]\int d^2x\partial_\nu(\eta_++\eta_-)^2
=-\frac{\hbar^2}{2M}A^2\Big[{1\over z} {\partial \nu\over \partial z} +\frac{\partial^2\nu}{\partial z^2}\Big]\partial_\nu\frac1{A^2}\nonumber\\
&=&-\frac{\hbar^2}{2M}\Big[{1\over z} {\partial \nu\over \partial z} +\frac{\partial^2\nu}{\partial z^2}\Big](A^2\partial_\nu A^{-2})
.\end{eqnarray}
Again this is an exact computation. Let us now find the leading contribution of this expression, we write below  the expansion of $A^2\partial_\nu A^{-2}$ to leading order,
\begin{eqnarray}
A^2\partial_\nu A^{-2} 
&\approx& -{2\over \nu }+...
,\end{eqnarray}
the details of the calculation are given in an Appendix, using our leading order expansions for ${\partial \nu\over  \partial z}$ and ${\partial^2 \nu \over \partial z^2}$ given by (\ref{nufirstder}) and (\ref{nusecondder}) respectively, as well as the expansion above, we find,
\begin{eqnarray}
(1_c)\approx\frac{\hbar^2}{M}\frac{1}{4z^2}+...
\end{eqnarray}
In a similar way, the last part $(1_d)$ is computed,
\begin{eqnarray}
-\frac{\hbar^2}{M}A^2\left(\frac{\partial\nu}{\partial z}\right)^2\int d^2x(\eta_++\eta_-)\partial_\nu^2(\eta_++\eta_-)&\!\!\!\!\!\!\!\!\!=&\!\!\!\!\!\!\!\!\!-\frac{\hbar^2}{M}A^2\left(\frac{\partial\nu}{\partial z}\right)^2\int d^2x\left[\frac1{2}\partial_\nu^2(\eta_++\eta_-)^2-(\partial_\nu(\eta_++\eta_-))^2\right]\nonumber\\
&=&-\frac{\hbar^2}{M}A^2\left(\frac{\partial\nu}{\partial z}\right)^2\left[\frac1{2}\partial_\nu^2\frac1{A^2}-\int d^2x(\partial_\nu(\eta_++\eta_-))^2\right]\nonumber
.\end{eqnarray}
Up to this point there is no approximation, we calculate the last part separately,
\begin{eqnarray}
\int d^2x (\partial_\nu(\eta_++\eta_-))^2 
&=&\frac{8\nu^2}{\hbar^2}\frac{\pi^2\hbar^2}{m^2}\int_0^\infty dt_1dt_2t_1t_2e^{-\frac{\nu^2}{\hbar}(t_1+t_2)}\Bigg[  K_{t_1+t_2}(0)+K_{t_1+t_2}(z/2,-z/2)\Bigg]\nonumber\\
&=&\frac{\nu^2}{\hbar^2}\frac{\pi^2\hbar^2}{m_*^2}\int_{-t}^t ds (t^2-s^2)\int_0^\infty dte^{-\frac{\nu^2}{\hbar}t}\Bigg[ K_t(0)+K_t(z/2, -z/2) \Bigg]\nonumber\\
&=&\frac{4\nu^2}{3\hbar^2}\frac{\pi^2\hbar^2}{m_*^2}\int_0^\infty dt t^3e^{-\frac{\nu^2}{\hbar}t}\Bigg[K_t(0)+K_t(z/2,-z/2) \Bigg]\nonumber\\
&=&\frac{4\nu^2}{3\hbar^2}\frac{\pi\hbar}{2m_*}\int_0^\infty dt t^2e^{-\frac{\nu^2}{\hbar}t}\Bigg[1+e^{-\frac{2m_*z^2}{4\hbar t}}\Bigg]\nonumber\\
&=&\frac{4\nu^2}{3\hbar^2}\frac{\pi\hbar}{2m_*}\Bigg[ {2\hbar^3\over \nu^6}+{\hbar^3\over 4 \nu^6} \Big({\sqrt{2m_*} \nu \over \hbar}z\Big)^3K_3\left({\sqrt{2m_*}\nu\over \hbar}z\right)\Bigg]
.\end{eqnarray}
Finally,
\begin{eqnarray}
\!\!\!\!(1_d)=-\frac{\hbar^2}{M}A^2\left(\frac{\partial\nu}{\partial z}\right)^2\left(\frac1{2}\partial_\nu^2\frac1{A^2}-\frac{4\nu^2}{3\hbar^2}\frac{\pi\hbar}{2m_*}\Bigg[ {2\hbar^3\over \nu^6}+{\hbar^3\over 4\nu^6} \Big({\sqrt{2m_*} \nu \over \hbar}z\Big)^3K_3\left({\sqrt{2m_*}\nu\over \hbar}z\right)\Bigg]\right)
.\end{eqnarray}
 As a result of a careful calculation, the details of which are given in an Appendix,  we find the leading order of $A^2 \partial_\nu^2 A^{-2}$ term, 
\begin{equation}
A^2\partial_\nu^2 A^{-2} \approx {6\over \nu^2}+...
\end{equation} 
After an expansion of all  these terms, again using the small argument behavior of Bessel functions, to the leading order, we find that,
\begin{eqnarray}
(1_d)&\approx&\frac{\hbar^2}{M}\left[-\frac{3}{4z^2}+\frac{2}{3z^2}\right]\approx-\frac{\hbar^2}{M}{1\over 12z^2}+...
\end{eqnarray}
As a result the leading term of the first part labeled as $(1)$ above, is found as,
\begin{equation}
(1)\approx {\hbar^2\over M} \left[ {1\over 4z^2}-{1\over 12 z^2}\right]+...
\end{equation}
The next term here is labeled as $(2)$, 
\begin{equation}
-\frac{\hbar^2}{M}\frac1{A}\nabla_z^2 A=-\frac{\hbar^2}{M}\frac1{A}\left(\frac{\partial^2}{\partial z^2}+\frac1{z}\frac{\partial}{\partial z}\right)A
\end{equation}
Therefore, we need these derivatives, we collect the calculations in an Appendix, and we  just state the result of these expansions.   The first derivative term, to the leading order, is
\begin{eqnarray}
\frac1{A}\left(\frac1{z}\partial_z A\right)\approx-\frac1{2z^2}+...
.\end{eqnarray}
The next derivative becomes, again to the leading order,
\begin{equation}
\frac1{A}\partial_z^2A\approx \frac{3}{4z^2}+...
\end{equation}
As a consequence, the total derivative of normalization can be expanded in leading order  as,
\begin{equation}
-{\hbar^2\over M}\frac1{A}\nabla_z^2 A\approx-{\hbar^2\over M}\frac1{4z^2}+...
.\end{equation}

There is one more term, which corresponds to the expression number $(3)$,
\begin{eqnarray}
(3)&=&-\frac{\hbar^2}{M}2A\partial_z A\int d^2x(\eta_++\eta_-)\left(\partial_z|_\nu(\eta_++\eta_-)+\frac{\partial\nu}{\partial z}\partial_\nu(\eta_++\eta_-)\right)\nonumber\\
&=&-\frac{\hbar^2}{M}2A\partial_z A\int d^2x\Bigg[\eta_+\partial_z|_\nu \eta_++\eta_-\partial_z|_\nu\eta_-+\eta_+\partial_z|_\nu\eta_-+\eta_-\partial_z|_\nu\eta_++\frac{\partial\nu}{\partial z}(\eta_++\eta_-)\partial_\nu(\eta_++\eta_-)\Bigg].\nonumber
\end{eqnarray}
To simplify our calculations, we make the following observation,
\begin{equation}
\partial_z|_\nu\int d^2x(\eta_+^2+\eta_-^2)=0.
\end{equation}
As a result,  equation $(3) $ becomes
\begin{equation}
(3)=-\frac{\hbar^2}{M}2A\partial_z A\int d^2x\left[\partial_z|_\nu\eta_+\eta_- +\frac1{2}\frac{\partial\nu}{\partial z}\partial_\nu(\eta_++\eta_-)^2\right].
\end{equation}
We divide the above term into two parts, the first of which can be easily found,
\begin{eqnarray}
(3_a)&=&-\frac{\hbar^2}{M}A\partial_z A\frac{\partial\nu}{\partial z}\partial_\nu\int d^2x (\eta_++\eta_-)^2=-\frac{\hbar^2}{M}A\partial_z A\frac{\partial\nu}{\partial z}\partial_\nu\frac1{A^2}=-\frac{\hbar^2}{M}{1\over A}\partial_z A\frac{\partial\nu}{\partial z}A^2\partial_\nu\frac1{A^2}\nonumber\\
&\approx&\frac{\hbar^2}{M}\frac1{2z^2}+...,
\end{eqnarray}
where in the last line we present the short distance expansion of this part. In a similar way, we work out the second part,
\begin{eqnarray}
(3_b)&=&-\frac{\hbar^2}{M}2A\partial_z A\partial_z|_\nu\int d^2x\eta_+\eta_-=-\frac{\hbar^2}{M}2A\partial_z A\frac{\pi\hbar}{2m_*}\partial_z|_\nu\left[\frac{\sqrt{2m_*}}{\nu}zK_1\left(\frac{\sqrt{2m_*}}{\hbar}\nu z\right)\right]
\end{eqnarray}
This expression is worked out in the Appendix,  its expansion  under the small $z/\zeta_0$ approximation leads to
\begin{eqnarray}
(3_b)&\approx&O\Big(\frac{m_*}{M}\nu^2\Big)+...,
\end{eqnarray}
hence it is of lower order.
Consequently the whole sum for $(3)$ becomes in the leading approximation;
\begin{equation}
(3)\approx {\hbar^2\over M} {1\over 2z^2}+... 
\end{equation}

We will now consider the cross term which contains first order derivative of the heavy particle wave functions in the averaged out Schr\"odinger equation given in equation (\ref{averaged}),  
\begin{equation}
(*)=-\frac{\hbar^2}{M}\int d^2x \phi(x|z)\nabla_z \phi(x|z)\cdot\nabla_z \Psi(z) 
.\end{equation}
We evaluate each term here, 
\begin{eqnarray}
(*)&= &-\frac{\hbar^2}{M}2\frac{\partial \Psi}{\partial z}\int d^2x A(\eta_++\eta_-)\left[(\eta_++\eta_-){\partial \over  \partial z}A+A{\partial \over \partial z}(\eta_++\eta_-)\right]\nonumber\\
&=&-\frac{\hbar^2}{M}\frac{\partial\Psi}{\partial z}\frac{2}{A}\partial_z A-\frac{\hbar^2}{M}\frac{\partial\Psi}{\partial z}2A^2\int d^2x(\eta_++\eta_-)\left(\partial_z|_\nu+\frac{\partial\nu}{\partial z}\partial_\nu\right)(\eta_++\eta_-)\nonumber\\
&= &-\frac{\hbar^2}{M}\frac{2}{A}\partial_z A\partial_z\Psi -\frac{\hbar^2}{M}\frac{\partial\Psi}{\partial z}2A^2\int d^2x\left(\frac1{2}\partial _z|_\nu(\eta_+^2+\eta_-^2)+\partial_ z|_\nu\eta_+\eta_- +\frac1{2}\frac{\partial\nu}{\partial z}\partial_\nu(\eta_++\eta_-)^2\right)\nonumber\\
&=&-\frac{\hbar^2}{M}\frac{2}{A}\partial_z A\partial_z\Psi -\frac{\hbar^2}{M}\partial_z\Psi 2A^2\left(\frac{\pi\hbar}{2m_*}{\partial\over \partial z}\Big|_\nu\frac{\sqrt{2m_*}}{\nu}|z|K_1\left(\frac{\sqrt{2m_*}}{\hbar}\nu z\right)+\frac1{2}\frac{\partial\nu}{\partial z}\partial_\nu\frac1{A^2}\right)\nonumber
.\end{eqnarray}
We now evaluate the derivative and use the Bessel function identity,
\begin{eqnarray}
 xK_{n-1}(x)-xK_{n+1}(x)=-2nK_n(x),
\end{eqnarray}
 for the resulting $K_2(\cdot)$, to arrive at,
\begin{eqnarray}
&\ & -\frac{\hbar^2}{M}\partial_z\Psi\Bigg({2\over A}\partial_z A +A^2\frac{\pi\hbar}{2m_*}\Bigg[\frac{\sqrt{2m_*}}{\nu}K_1\left(\frac{\sqrt{2m_*}}{\hbar}\nu z\right)-\frac{\sqrt{2m_*}}{\nu}z\frac{\sqrt{2m_*}\nu}{\hbar}\Big[\frac{\sqrt{2m_*}\nu}{\hbar}z\Big]^{-1}K_1\left(\frac{\sqrt{2m_*}}{\hbar}\nu z\right) \nonumber\\
&\ & \quad -\frac{\sqrt{2m_*}}{\nu}z\frac{\sqrt{2m_*}\nu}{\hbar}K_0\left(\frac{\sqrt{2m_*}}{\hbar}\nu z\right)\Bigg]+\frac{\partial\nu}{\partial z}A^2\partial_\nu\frac1{A^2}\Bigg)\nonumber\\
\end{eqnarray}
After cancelling $K_1(\cdot)$ terms, we expand this term to leading order, using the result for $A^2\partial_\nu A^{-2}$ that we found before, we  find,
\begin{eqnarray}
(*)&\approx& C_1\frac{m_*}{M}\nu^2 z\partial_z\Psi\ln\left(\frac{\sqrt{2m_*}}{\hbar}\nu z\right)+...
,\end{eqnarray}
where $C_1$ is a constant that we can explicitly compute.
This is quite a remarkable result, we see that there is no ${1\over z}$ term multiplying $\partial_z\Psi$ term. To get a well-defined operator we need to symmetrize this term, and the absence of ${1\over z}$ implies that such a symmetrization will not lead to another ${1\over z^2}$ correction to the potential term.

Finally, we write down all the leading terms for  the (effective) Schr\"odinger equation for the heavy degrees of freedom in the leading approximation,
\begin{eqnarray}
&\ & -\frac{\hbar^2}{M}{\nabla_z^2}\Psi(z)-\nu^2(|z|)\Psi(z)-\frac{\hbar^2}{M}\int d^2x \phi(x|z)\nabla_z \phi(x|z)\cdot\nabla_z \Psi(z) \nonumber\\
&\ &\quad\quad \quad  +\Bigg(\int d^2x \phi(x|z)\Big[-\frac{\hbar^2}{M}{\nabla_z^2}\Big]_{\rm reg}\phi(x|z)\Bigg) \Psi(z)\nonumber\\
&\ & \quad \approx -\frac{\hbar^2}{M}{\nabla_z^2}\Psi(z)-\Big[{\hbar\over \sqrt{2m_*}}{2\epsilon\over z e^\gamma}+...\Big]\Psi(z)+\Big[{\hbar^2\over M} {5\over 12 z^2}+...\Big]\Psi(z)+...=\delta E\Psi(z).\nonumber
\end{eqnarray}
This is a very interesting result, since the leading energy is not only given by the potential term that we have in the Born-Oppenheimer approximation but gets a nonperturbative contribution from  the effective potential generated by the kinetic energy operator acting on the wave function ofthe light degrees of freedom. The energy that one finds is not small by any means, it grows with the mass ratio of the heavy one to the one of the light particle, namely $M/m$. 
Nevertheless, it  has still an order by order expansion, the real expansion is being done with respect to the smallness of the average separation of the heavy particles relative to the spread of the light particle's wave function. The solution of the above equation is well-known in terms of the generalized Laguerre polynomials, being familiar from the hydrogenic atoms solutions. In the ground state wave function we have no dependence on the angular coordinate therefore no angular momentum contribution. We write the resulting equation
\begin{equation}
-\frac{\hbar^2}{M}\left[\frac{\partial^2}{\partial z^2}+\frac{1}{z}\frac{\partial}{\partial z}-\frac{\beta^2}{z^2}\right]\Psi(z)-\frac{\alpha}{z}\Psi(z)=\delta E\Psi(z)
\end{equation}
by defining the following  parameters and the new coordinate $r$,
\begin{eqnarray}
\delta E=-\frac{E_0}{K^2(n)}\qquad {\rm where} \qquad E_0=\frac{\alpha^2 M}{4\hbar^2}\qquad{\rm also}\quad \beta^2=5/12\nonumber\\
r=\frac{z}{z_0K(n)}\qquad{\rm where}\qquad z_0=\frac{\hbar^2}{M\alpha}\qquad {\rm and}\qquad \alpha=\frac{2\hbar}{\sqrt{2m_*}}{\epsilon\over e^\gamma}
\end{eqnarray}
we have a transformed equation:
\begin{equation}
\left[\frac{\partial^2}{\partial r^2}+\frac{1}{r}\frac{\partial}{\partial r}-\frac{(\beta^2)}{r^2}+\frac{K}{r}-\frac1{4}\right]R_{n}=0.
\end{equation}
Considering $ r \rightarrow \infty $ limit, one can see that we can write the solution in the form,
\begin{equation}
R_{n}=C(n)e^{-\frac{r}{2}}\psi(r),
\end{equation}
and putting this back into the equation, we have  for $\psi(r) $,
\begin{equation}
\frac{\partial^2\psi}{\partial r^2}+\left(\frac{1}{r}-1\right)\frac{\partial\psi}{\partial r}+\frac1{r}\left(K-\frac1{2}\right)\psi-\frac{\beta^2}{r^2}\psi=0
.\end{equation}
We investigate small $r$ behavior of the equation, and look for a solution by putting  $r^\delta $ into the  most singular part of the  equation (for $r\to 0^+$),
$$
\delta(\delta-1)r^{\delta-2}+\frac1{r}\delta r^{\delta-1}-(\beta^2)r^{\delta-2}=0,
$$
we can easily see that
\begin{equation}
\delta^2=\beta^2
\end{equation}
and we admit the regular solution, $\delta=\beta$, so the wave function $\psi(r)$ should be of the form
\begin{equation}
\psi(r)=r^{\beta}g(r),
\end{equation}
for a regular function $g(r)$. Let's put this back into the equation, to get 
\begin{equation}
r\frac{\partial^2g(r)}{\partial r^2}+(2\beta+1-r)\frac{\partial g(r)}{\partial r}+\left(K-\frac1{2}-\beta\right)g(r)=0.
\end{equation}
A general soluiton which is regular at $r=0$ of this equation is
\begin{equation}
g(r)=L_{K-\frac1{2}-\beta}^{2\beta}(r)
.\end{equation}
If we place no restrictions on the parameters the Laguerre functions grow exponentially at infinity, the normalizable wave function assumption puts a restriction on the allowed values as,
$$
K(n)-\frac1{2}-\beta= n
,$$
for any integer $n\geq 0$, thus the resultant wave function becomes
\begin{equation}
R_{n}=C(n)r^{\beta}e^{-\frac{r}{2}}L_{n}^{2\beta}(r)
.\end{equation}
Consequently the energy becomes,
$$
E_n=-\frac{\alpha^2M}{4\hbar^2}\frac1{\left(n+\frac1{2}+\beta\right)^2}=-\frac{M}{2m_*}\frac{\epsilon^2}{e^{2\gamma}\left(n+\frac1{2}+\beta\right)^2}
$$
this gives for the true ground state for $n=0$,
$$
E_g=-\frac{M}{2m}\frac{e^{-2\gamma}\epsilon^2}{(1+2\beta)^2}.
,$$
where we replaced $m_*$ with $m$ to have the correct leading order expression.
Moreover, we have the ground state wave function
$$
R_0=Cr^\beta e^{-\frac{r}{2}}\qquad {\rm and\ here }\qquad   r=\frac{2z}{z_0(1+2\beta)}
.$$
These are our main results, it is essential to note that the energy goes with $M/m$ ratio, this is important because as we let the heavy particle mass becomes infinite, the two heavy centers coalesce into a single center, renormalization that we perform does not allow this configuration, hence we find a divergence. A crucial point to check for the consistency is the expectation value of the separation $z$ {\it within our approximation}, that is with the solution we have found for the heavy particle separation. Not surprisingly we find,
\begin{eqnarray}
<z>&=&\int_0^\infty \int_0^{2\pi} z dz d\theta |\Psi(z)|^2 z\nonumber\\
&=&\frac{e^\gamma}{4}\frac{\Gamma(2\beta+3)}{\Gamma(2\beta+2)}(1+2\beta)\frac{\hbar}{\sqrt{2m_*}\epsilon}\left(\frac{2m_*}{M}\right)\nonumber\\
&=&\frac{e^\gamma}{2}\frac{\Gamma(2\beta+3)}{\Gamma(2\beta+2)}(1+2\beta)\left(\frac{m_*}{M}\right)\zeta_0
,\end{eqnarray}
where $\zeta_0$ refers to  the charateristic length scale of the light particle wave function.

 In the last Appendix we compute the next order corrections to the energy coming from the higher order expansions. These are not computed completely but some terms are found to illustrate the consistency of our approximations.

\section{Conclusions}
In a singular system which has two heavy and one light particle, we can apply the Born-Oppenheimer approximation under the assumption that the spread of the wave function of the light particle is much much larger than the spread of the  heavy particles relative coordinate wave function, here the relative coordinate is characterized by the variable $z$.
Contrary to the usual Born-Oppenheimer expansions, we find that the contributions of the heavy degrees of freedom to the total energy of the system is very large, of order $M/m$. Yet there is still an order by order expansion of the total energy of the system in terms of the expectation values of increasing powers of $z$ and logarithms of $z$. 
It would be very important to understand this from a many body perspective as we have done in our previous work for a one dimensional version of this problem. We plan to investigate this approach in the near future.
\section{Acknowledgements} O. T. Turgut would like to thank F. Erman, J. Hoppe, A. Moustafazadeh, M. Znojil for discussions. Part of this work is completed while the second author is visiting Mathematics Department of KTH, Stockholm, he is grateful to J. Hoppe for his extermely  kind invitation to work there.    

\section{Appendix-Small Distance Expansions}
Here, we provide the detailed computations of all the derivative terms and their expansions. Short distance expansions are used to find the leading order solution within this modified Born-Oppenheimer approximation. For all the approximations below we make use of the Bessel function expansion for $n\geq 2$,
\begin{equation}
K_n(x)\approx {1\over 2}(n-1)! {1\over (x/2)^n}+...
\end{equation}
for small $x$ and also the expansion of $K_0(x)$ that we mentioned in the text. Moreover in some cases $K_1(x)$'s next order term in the expansion may have importance and it is given by 
\begin{equation}
K_1(x)\approx {1\over 2} {1\over (x/2) }+{1\over 2} x \ln({x\over 2})+...\quad {\rm as} \ x\to 0^+
,\end{equation}

In our calculations we need the second derivative of the square-root of the binding energy $\nu(z)$,
\begin{eqnarray}
\frac{\partial^2\nu}{\partial z^2}&=&\!\!-2\frac{\frac{\sqrt{2m_*}}{\hbar}\nu\frac{\partial\nu}{\partial z}K_1\left(\frac{\sqrt{2m_*}}{\hbar}\nu z\right)}{\left[1+\frac{\sqrt{2m_*}}{\hbar}\nu zK_1\left(\frac{\sqrt{2m_*}}{\hbar}\nu z\right)\right]}+\frac{\frac{2m_*}{\hbar^2}\nu^2\left(z\frac{\partial\nu}{\partial z}+\nu\right)\left[K_0\left(\frac{\sqrt{2m_*}}{\hbar}\nu z\right)+\frac{\hbar}{\sqrt{2m_*}\nu z}K_1\left(\frac{\sqrt{2m_*}}{\hbar}\nu z\right)\right]}{\left[1+\frac{\sqrt{2m_*}}{\hbar}\nu zK_1\left(\frac{\sqrt{2m_*}}{\hbar}\nu z\right)\right]}\nonumber\\
&+&\frac{\frac{2m_*}{\hbar^2}\nu^2\left(z\frac{\partial\nu}{\partial z}+\nu\right)\left(K_1\left(\frac{\sqrt{2m_*}}{\hbar}\nu z\right)\right)^2}{\left[1+\frac{\sqrt{2m_*}}{\hbar}\nu zK_1\left(\frac{\sqrt{2m_*}}{\hbar}\nu z\right)\right]^2}\nonumber\\
&-&\frac{\frac{(2m_*)^\frac{3}{2}}{\hbar^3}\nu^3|z|\left(z\frac{\partial\nu}{\partial z}+\nu\right)K_1\left(\frac{\sqrt{2m_*}}{\hbar}\nu z\right)}{\left[1+\frac{\sqrt{2m_*}}{\hbar}\nu zK_1\left(\frac{\sqrt{2m_*}}{\hbar}\nu z\right)\right]^2}\left[K_0\left(\frac{\sqrt{2m_*}}{\hbar}\nu z\right)+\frac{\hbar}{\sqrt{2m_*}\nu z}K_1\left(\frac{\sqrt{2m_*}}{\hbar}\nu z\right)\right]\nonumber
\\
&=&\!\!-2\frac{\frac{\sqrt{2m_*}}{\hbar}\nu\frac{\partial\nu}{\partial z}K_1\left(\frac{\sqrt{2m_*}}{\hbar}\nu z\right)}{\left[1+\frac{\sqrt{2m_*}}{\hbar}\nu zK_1\left(\frac{\sqrt{2m_*}}{\hbar}\nu z\right)\right]}+\frac{\frac{2m_*}{\hbar^2}\nu^2\left(z\frac{\partial\nu}{\partial z}+\nu\right)\left[K_0\left(\frac{\sqrt{2m_*}}{\hbar}\nu z\right)+\frac{\hbar}{\sqrt{2m_*}\nu z}K_1\left(\frac{\sqrt{2m_*}}{\hbar}\nu z\right)\right]}{\left[1+\frac{\sqrt{2m_*}}{\hbar}\nu zK_1\left(\frac{\sqrt{2m_*}}{\hbar}\nu z\right)\right]}\nonumber
\end{eqnarray}
\begin{eqnarray}
&-&\frac{\frac{(2m_*)^\frac{3}{2}}{\hbar^3}\nu^3z\left(z\frac{\partial\nu}{\partial z}+\nu\right)K_1\left(\frac{\sqrt{2m_*}}{\hbar}\nu z\right)}{\left[1+\frac{\sqrt{2m_*}}{\hbar}\nu zK_1\left(\frac{\sqrt{2m_*}}{\hbar}\nu z\right)\right]^2}K_0\left(\frac{\sqrt{2m_*}}{\hbar}\nu z\right)\nonumber\\
&=&-\frac{\frac{\sqrt{2m_*}}{\hbar}\nu\frac{\partial\nu}{\partial z}K_1\left(\frac{\sqrt{2m_*}}{\hbar}\nu z\right)}{\left[1+\frac{\sqrt{2m_*}}{\hbar}\nu z K_1\left(\frac{\sqrt{2m_*}}{\hbar}\nu z\right)\right]}-\frac1{z}\frac{\partial\nu}{\partial z}\nonumber\\
&+&\frac{\frac{2m_*}{\hbar^2}\nu^2\left(z\frac{\partial \nu}{\partial z}+\nu\right)K_0\left(\frac{\sqrt{2m_*}}{\hbar}\nu z\right)}{\left[1+\frac{\sqrt{2m_*}}{\hbar}\nu zK_1\left(\frac{\sqrt{2m_*}}{\hbar}\nu z\right)\right]}\left[1-\frac{\frac{\sqrt{2m_*}}{\hbar}\nu zK_1\left(\frac{\sqrt{2m_*}}{\hbar}\nu z\right)}{\left[1+\frac{\sqrt{2m_*}}{\hbar}\nu zK_1\left(\frac{\sqrt{2m_*}}{\hbar}\nu z\right)\right]}\right]\nonumber
\end{eqnarray}
\begin{eqnarray}
&=&-\frac1{2}\frac{\sqrt{2m_*}}{\hbar}\nu\left[-\frac{\nu}{2z}-\frac1{8}\frac{2m_*}{\hbar^2}\nu^3z\ln\left(\frac{\sqrt{2m_*}}{2\hbar}\nu z\right)\right]\left[\frac{\hbar}{\sqrt{2m_*}\nu z}+\frac1{4}\frac{\sqrt{2m_*}}{\hbar}\nu z\ln\left(\frac{\sqrt{2m_*}}{2\hbar}\nu z\right)\right]\nonumber\\
&+&\frac{\nu}{2z^2}+\frac1{8}\frac{2m_*}{\hbar^2}\nu^3\ln\left(\frac{\sqrt{2m_*}}{2\hbar}\nu z\right)\nonumber\\
&+&\frac1{2}\frac{2m_*}{\hbar^2}\nu^2\left[\frac{\nu}{2}-\frac1{8}\frac{2m_*}{\hbar^2}\nu^3z^2\ln\left(\frac{\sqrt{2m_*}}{2\hbar}\nu z\right)\right]\left[1-\frac1{4}\frac{2m_*}{\hbar^2}\nu^2z^2\ln\left(\frac{\sqrt{2m_*}}{2\hbar}\nu z\right)\right]\nonumber\\
&\ &\qquad\qquad\times\left[-\ln\left(\frac{\sqrt{2m_*}}{2\hbar}\nu z\right)-\gamma\right]\left[\frac1{2}-\frac1{8}\frac{2m_*}{\hbar^2}\nu^2z^2\ln\left(\frac{\sqrt{2m_*}}{2\hbar}\nu z\right)\right]\nonumber\\
&=&\frac{3}{4}\frac{\nu}{z^2}+a_1\frac{2m_*}{\hbar^2}\nu^3+a_2\frac{2m_*}{\hbar^2}\nu^3\ln\left(\frac{\sqrt{2m_*}}{2\hbar}\nu z\right)+...\nonumber
.\end{eqnarray}
As a result we now see that, 
\begin{eqnarray}
\frac1{\nu}\frac{\partial^2\nu}{\partial z^2}&=&\frac{3}{4z^2}+O\left(\frac{2m_*}{\hbar^2}\nu^2\right)+O\left(\frac{2m_*}{\hbar^2}\nu^2\ln\left(\frac{\sqrt{2m_*}}{2\hbar}\nu|z|\right)\right),
\end{eqnarray}
here $a_1,a_2$ are constants that we can find explicitly, moreover we did not further simplify the $\nu$ terms to see the pattern more explicitly, otherwise they should also be expanded.

In our computations we also need various derivatives of the normalization constant $A$. We will now start with the derivative of the inverse with respect to $\nu$ only keeping $z$ constant,
\begin{eqnarray}
\partial_\nu A^{-2}&=&2\pi\frac{\hbar^2}{2m_*}\Bigg[-\frac{2}{\nu^3}\left[1+\frac{\sqrt{2m_*}}{\hbar}\nu zK_1\left(\frac{\sqrt{2m_*}}{\hbar}\nu z\right)\right]+\frac1{\nu^2}\frac{\sqrt{2m_*}}{\hbar}zK_1\left(\frac{\sqrt{2m_*}}{\hbar}\nu z\right)\nonumber\\
&-&\frac1{\nu}\frac{2m_*}{\hbar^2}z^2\left[K_0\left(\frac{\sqrt{2m_*}}{\hbar}\nu z\right)+\frac{\hbar}{\sqrt{2m_*}\nu z}K_1\left(\frac{\sqrt{2m_*}}{\hbar}\nu z\right)\right]\Bigg]\nonumber\\
&=&2\pi\frac{\hbar^2}{2m_*}\Bigg[-\frac{2}{\nu^3}\left[1+\frac{\sqrt{2m_*}}{\hbar}\nu zK_1\left(\frac{\sqrt{2m_*}}{\hbar}\nu z\right)\right]-\frac1{\nu}\frac{2m_*}{\hbar^2}z^2K_0\left(\frac{\sqrt{2m_*}}{\hbar}\nu z\right)\Bigg]\nonumber
\end{eqnarray}
Let us now write the expression we need, and then expand for small $z$;
\begin{eqnarray}
A^2\partial_\nu A^{-2}&=& -\frac{2}{\nu}-\nu z^2\frac{2m_*}{\hbar^2}\frac{K_0\left(\frac{\sqrt{2m_*}}{\hbar}\nu|z|\right)}{\left(1+\frac{\sqrt{2m_*}}{\hbar}\nu zK_1\left(\frac{\sqrt{2m_*}}{\hbar}\nu z\right)\right)}\nonumber\\
&\approx&-\frac{2}{\nu}-\frac1{2}\nu z^2\frac{2m_*}{\hbar^2}\left[-\ln\left(\frac{\sqrt{2m_*}}{2\hbar}\nu z\right)-\gamma\right]\left[1-\frac1{4}\frac{2m_*}{\hbar^2}\nu^2z^2\ln\left(\frac{\sqrt{2m_*}}{2\hbar}\nu z\right)+...\right]\nonumber\\
&\approx&-\frac{2}{\nu}+b_1\frac{2m_*}{\hbar^2}\nu z^2+b_2\frac{2m_*}{\hbar^2}\nu z^2\ln\left(\frac{\sqrt{2m_*}}{2\hbar}\nu z\right)+...
,\end{eqnarray}
here $b_1, b_2$ again are some constants we can determine, furthermore, we did not expand the $\nu$ terms, in principle they should also be expanded, however this form is better since in our computations some combinations of $\nu$'s will then cancel to give us a simpler result.

Let us now consider the following second $\nu$ derivative of $A^{-2}$ that we need in our calculations:
\begin{eqnarray}
\partial^2_\nu A^{-2}&=& 2\pi\frac{\hbar^2}{2m_*}\Bigg[\frac{6}{\nu^4}\left[1+\frac{\sqrt{2m_*}}{\hbar}\nu zK_1\left(\frac{\sqrt{2m_*}}{\hbar}\nu z\right)\right]-\frac{2}{\nu^3}\frac{\sqrt{2m_*}}{\hbar}zK_1\left(\frac{\sqrt{2m_*}}{\hbar}\nu z\right)\nonumber\\
&+&\frac{2}{\nu^2}\frac{2m_*}{\hbar^2}z^2\left[K_0\left(\frac{\sqrt{2m_*}}{\hbar}\nu z\right)+\frac{\hbar}{\sqrt{2m_*}\nu z}K_1\left(\frac{\sqrt{2m_*}}{\hbar}\nu z\right)\right]\nonumber\\
&+&\frac1{\nu^2}\frac{2m_*}{\hbar^2}z^2K_0\left(\frac{\sqrt{2m_*}}{\hbar}\nu z\right)+\frac1{\nu}\frac{(2m_*)^\frac{3}{2}}{\hbar^3} z^3K_1\left(\frac{\sqrt{2m_*}}{\hbar}\nu z\right)\Bigg]\nonumber
\end{eqnarray}
We now write the combination we need and then expand as usual keeping $\nu$ terms as they are, so as to simplify later calculations, 
\begin{eqnarray}
A^2\partial^2_\nu A^{-2}&=&\frac{6}{\nu^2}+3z^2\frac{2m_*}{\hbar^2}\frac{K_0\left(\frac{\sqrt{2m_*}}{\hbar}\nu z\right)}{\left[1+\frac{\sqrt{2m_*}}{\hbar}\nu zK_1\left(\frac{\sqrt{2m_*}}{\hbar}\nu z\right)\right]}+\nu z^3\frac{(2m_*)^\frac{3}{2}}{\hbar^3}\frac{K_1\left(\frac{\sqrt{2m_*}}{\hbar}\nu z\right)}{\left[1+\frac{\sqrt{2m_*}}{\hbar}\nu zK_1\left(\frac{\sqrt{2m_*}}{\hbar}\nu z\right)\right]}\nonumber\\
&\approx&\!\!\!\frac{6}{\nu^2}+\frac{3}{2}z^2\frac{2m_*}{\hbar^2}\left[-\ln\left(\frac{\sqrt{2m_*}}{2\hbar}\nu z\right)-\gamma\right]\left[1-\frac1{4}\frac{2m_*}{\hbar^2}\nu^2z^2\ln\left(\frac{\sqrt{2m_*}}{2\hbar}\nu z\right)\right]\nonumber\\
&+&\!\!\!\frac1{2}\nu z^3\frac{(2m_*)^\frac{3}{2}}{\hbar^3}\left[\frac{\hbar}{\sqrt{2m_*}\nu z}+\frac1{2}\frac{\sqrt{2m_*}}{\hbar}\nu z\ln\left(\frac{\sqrt{2m_*}}{2\hbar}\nu z\right)\right]\left[1-\frac1{4}\frac{2m_*}{\hbar^2}\nu^2z^2\ln\left(\frac{\sqrt{2m_*}}{2\hbar}\nu z\right)\right]\nonumber\\
&\approx&\!\!\!\frac{6}{\nu^2}+c_1 z^2\frac{2m_*}{\hbar^2}+c_2z^2\frac{2m_*}{\hbar^2}\ln\left(\frac{\sqrt{2m_*}}{2\hbar}\nu z\right)+...
,\end{eqnarray}
where we have well-defined constants $c_1, c_2$, and $\nu$ is kept as it is without further expansion. 
We now write down the combinations we encounter in our computations, one of them is this expression:
\begin{eqnarray}
&\ &-\frac{\hbar^2}{M}\left(\frac{\partial\nu}{\partial z}\right)^2\frac1{2}A^2\partial^2_\nu A^{-2}\nonumber\\
&\approx&-\frac{\hbar^2}{2M}\left[\frac{\nu^2}{4z^2}+\frac1{8}\frac{2m_*}{\hbar^2}\nu^4\ln\left(\frac{\sqrt{2m_*}}{2\hbar}\nu z\right)+...\right]\left[\frac{6}{\nu^2}+\frac{2m_*}{\hbar^2}z^2+\frac{2m_*}{\hbar^2}z^2\ln\left(\frac{\sqrt{2m_*}}{2\hbar}\nu z\right)+...\right]\nonumber\\
&\approx&-\frac{\hbar^2}{M}\frac{3}{4z^2}+O\left(\frac{m_*}{M}\nu^2\right)+O\left(\frac{m_*}{M}\nu^2\ln\left(\frac{\sqrt{2m_*}}{2\hbar}\nu z \right)\right)+...
\end{eqnarray}
We write the final expanded version that we use. Note that here instead of the constants in front we indicate the order of magnitude of the term in the expansion with big-$O$ symbol.
We also need the combination below, hence we write it explicitly and use the expansions we have obtained,
\begin{eqnarray}
&\ &-\frac{\hbar^2}{2M}\left[\frac1{z}\frac{\partial\nu}{\partial z}+\frac{\partial^2\nu}{\partial z^2}\right]A^2\partial_\nu A^{-2}\nonumber\\
&\approx&-\frac{\hbar^2}{2M}\left[-\frac{\nu}{2z^2}-\frac1{8}\frac{2m_*}{\hbar^2}\nu^3\ln\left(\frac{\sqrt{2m_*}}{2\hbar}\nu z\right)+\frac{3\nu}{4z^2}+\frac{2m_*}{\hbar^2}\nu^3+\frac{2m_*}{\hbar^2}\nu^3\ln\left(\frac{\sqrt{2m_*}}{2\hbar}\nu z\right)+...\right]\nonumber\\
&\ &\qquad\qquad\qquad\qquad\times\left[-\frac{2}{\nu}+\frac{2m_*}{\hbar^2}\nu z^2+\frac{2m_*}{\hbar^2}\nu z^2\ln\left(\frac{\sqrt{2m_*}}{2\hbar}\nu z\right)+...\right]\nonumber\\
&\approx&\frac{\hbar^2}{M}\frac1{4z^2}+O\left(\frac{m_*}{M}\nu^2\right)+O\left(\frac{m_*}{M}\nu^2\ln\left(\frac{\sqrt{2m_*}}{2\hbar}\nu z\right)\right)+...
\end{eqnarray}
Note that  the logarithmic terms by themselves are ambigous, one should keep the next order term to find the exact expansion, however we will not be able to compute all the corrections even at the first order due to complicated nature of these expressions, this is why we are essentially emphasizing the order of each term rather than the precise numerical factors.

Our second  purpose is to find derivatives of $A$, with respect to $z$,  to begin with, we  look at the first derivative of this normalization constant;
\begin{eqnarray}
&\ &\!\!\!\!\!\!\!\!\!\!\!\!\partial_z A=\frac1{\sqrt{2\pi}}\frac{\sqrt{2m_*}}{\hbar}\frac{\partial\nu}{\partial z}\frac1{\sqrt{1+\frac{\sqrt{2m_*}}{\hbar}\nu z K_1\left(\frac{\sqrt{2m_*}}{\hbar}\nu z\right)}}
-\frac1{2\sqrt{2\pi}}\frac{\sqrt{2m_*}}{\hbar}\nu\left[\frac{\frac{\sqrt{2m_*}}{\hbar}\left(z\frac{\partial\nu}{\partial z}+\nu\right)K_1\left(\frac{\sqrt{2m_*}}{\hbar}\nu z\right)}{\left(1+\frac{\sqrt{2m_*}}{\hbar}\nu zK_1\left(\frac{\sqrt{2m_*}}{\hbar}\nu z\right)\right)^\frac{3}{2}}\right]\nonumber\\
&\ &+\frac1{4\sqrt{2\pi}}\frac{\sqrt{2m_*}}{\hbar}\nu\left[\frac{\frac{2m_*}{\hbar^2}\nu z\left(z\frac{\partial\nu}{\partial z}+\nu\right)}{\left(1+\frac{\sqrt{2m_*}}{\hbar}\nu zK_1\left(\frac{\sqrt{2m_*}}{\hbar}\nu z\right)\right)^\frac{3}{2}}\right]\left[K_0\left(\frac{\sqrt{2m_*}}{\hbar}\nu z\right)+K_2\left(\frac{\sqrt{2m_*}}{\hbar}\nu z\right)\right].
\end{eqnarray}
We consider the following term
\begin{eqnarray}
&\ &\frac1{4\sqrt{2\pi}}\frac{\sqrt{2m_*}}{\hbar}\nu\left[\frac{\frac{2m_*}{\hbar^2}\nu z\left(z\frac{\partial\nu}{\partial z}+\nu\right)}{\left(1+\frac{\sqrt{2m_*}}{\hbar}\nu zK_1\left(\frac{\sqrt{2m_*}}{\hbar}\nu z\right)\right)^\frac{3}{2}}\right]\left[K_0\left(\frac{\sqrt{2m_*}}{\hbar}\nu z\right)+K_2\left(\frac{\sqrt{2m_*}}{\hbar}\nu z\right)\right]\nonumber\\
&=&\frac1{2\sqrt{2\pi}}\frac{\sqrt{2m_*}}{\hbar}\nu\left[\frac{\frac{2m_*}{\hbar^2}\nu z\left(z\frac{\partial\nu}{\partial z}+\nu\right)}{\left(1+\frac{\sqrt{2m_*}}{\hbar}\nu zK_1\left(\frac{\sqrt{2m_*}}{\hbar}\nu z\right)\right)^\frac{3}{2}}\right]\left[K_0\left(\frac{\sqrt{2m_*}}{\hbar}\nu z\right)+\frac{\hbar}{\sqrt{2m_*}\nu z}K_1\left(\frac{\sqrt{2m_*}}{\hbar}\nu z\right)\right]\nonumber\\
\end{eqnarray}
here we recognized a term corresponding to ${\partial \nu\over \partial z}$.
As a result of these computations, the first derivative of normalization constant is found to be, 
\begin{eqnarray}
\partial_z A&=&\frac1{\sqrt{2\pi}}\frac{\sqrt{2m_*}}{\hbar}\frac{\partial\nu}{\partial z}\frac1{\sqrt{1+\frac{\sqrt{2m_*}}{\hbar}\nu z K_1\left(\frac{\sqrt{2m_*}}{\hbar}\nu z\right)}}\nonumber\\
&+&\frac1{2\sqrt{2\pi}}\frac{\sqrt{2m_*}}{\hbar}\nu\left[\frac{\frac{2m_*}{\hbar^2}\nu|z|\left(z\frac{\partial\nu}{\partial z}+\nu\right)}{\left(1+\frac{\sqrt{2m_*}}{\hbar}\nu zK_1\left(\frac{\sqrt{2m_*}}{\hbar}\nu z\right)\right)^\frac{3}{2}}\right]K_0\left(\frac{\sqrt{2m_*}}{\hbar}\nu z\right)
.\end{eqnarray}
The term we are interested in is given below, by using the exact expression for $A$ again, we get,
\begin{eqnarray}
\frac1{A}\left(\frac1{z}\partial_z A\right)&=&\frac1{z\nu}\frac{\partial\nu}{\partial z}+\frac1{2z}\left[\frac{\frac{2m_*}{\hbar^2}\nu z\left(z\frac{\partial\nu}{\partial z}+\nu\right)}{\left(1+\frac{\sqrt{2m_*}}{\hbar}\nu zK_1\left(\frac{\sqrt{2m_*}}{\hbar}\nu z\right)\right)}\right]K_0\left(\frac{\sqrt{2m_*}}{\hbar}\nu z\right)\nonumber
.\end{eqnarray}
Let us consider its small $z$ expansion,  we therefore find,
\begin{eqnarray}
&\ &\frac1{A}\left(\frac1{z}\frac{\partial A}{\partial z}\right)=\frac1{\nu z}\frac{\partial\nu}{\partial z}+\frac1{2z}\left[\frac{\frac{2m_*}{\hbar^2}\nu z \left(z\frac{\partial\nu}{\partial z}+\nu\right)}{\left(1+\frac{\sqrt{2m_*}}{\hbar}\nu zK_1\left(\frac{\sqrt{2m_*}}{\hbar}\nu z\right)\right)}\right]K_0\left(\frac{\sqrt{2m_*}}{\hbar}\nu z\right)\nonumber\\
&\approx &\frac1{\nu z}\left[-\frac{\nu}{2 z}-\frac1{8}\frac{2m_*}{\hbar^2}\nu^3 z\ln\left(\frac{\sqrt{2m_*}}{2\hbar}\nu z\right)\right]\nonumber\\
&+&\frac1{4}\frac{2m_*}{\hbar^2}\nu\left[\frac{\nu}{2}-\frac1{8}\frac{2m_*}{\hbar^2}\nu^3z^2\ln\left(\frac{\sqrt{2m_*}}{2\hbar}\nu z\right)\right]\left[1-\frac1{4}\frac{2m_*}{\hbar^2}\nu^2z^2\ln\left(\frac{\sqrt{2m_*}}{2\hbar}\nu z\right)\right]\nonumber\\
&\ &\qquad\qquad\qquad\qquad\qquad\qquad\qquad\times\left[-\ln\left(\frac{\sqrt{2m_*}}{2\hbar}\nu z\right)-\gamma\right]\nonumber\\
&\approx&-\frac1{2z^2}+b_1\left(\frac{2m_*}{\hbar^2}\nu^2\ln\left(\frac{\sqrt{2m_*}}{2\hbar}\nu z\right)\right)+b_2\left(\frac{2m_*}{\hbar^2}\nu^2\right)+...,\nonumber
\end{eqnarray}
where the coefficients $b_1,b_2$ can be calculated from above, since we will not compute all the higher order corrections their precise values are not important. We may therefore write the required expansion as,
\begin{eqnarray}
&\ &-\frac{\hbar^2}{M}\frac1{A}\left(\frac1{z}\frac{\partial A}{\partial z}\right)\approx\frac{\hbar^2}{M}\frac1{2z^2}+O\left(\frac{m_*}{M}\nu^2\ln\left(\frac{\sqrt{2m_*}}{2\hbar}\nu z\right)\right)+O\left(\frac{m_*}{M}\nu^2\right)+...
\end{eqnarray}
In the last part, we  look at the more complicated derivative expressions of $A$;
\begin{eqnarray}
\partial_z A^2&=&\frac1{2\pi}\frac{2m_*}{\hbar^2}2\nu\frac{\partial\nu}{\partial z}\frac1{\left[1+\frac{\sqrt{2m_*}}{\hbar}\nu z K_1\left(\frac{\sqrt{2m_*}}{\hbar}\nu z\right)\right]}-\frac1{2\pi}\frac{(2m_*)^\frac{3}{2}}{\hbar^3}\nu^2\frac{\left( z\frac{\partial\nu}{\partial z}+\nu\right)K_1\left(\frac{\sqrt{2m_*}}{\hbar}\nu z\right)}{\left[1+\frac{\sqrt{2m_*}}{\hbar}\nu zK_1\left(\frac{\sqrt{2m_*}}{\hbar}\nu z\right)\right]^2}\nonumber\\
&+&\frac1{2\pi}\frac{(2m_*)^2}{\hbar^4}\nu^3 z\frac{\left(z\frac{\partial\nu}{\partial z}+\nu\right)}{\left[1+\frac{\sqrt{2m_*}}{\hbar}\nu z K_1\left(\frac{\sqrt{2m_*}}{\hbar}\nu z\right)\right]^2}\left[K_0\left(\frac{\sqrt{2m_*}}{\hbar}\nu z\right)+\frac{\hbar}{\sqrt{2m_*}\nu z}K_1\left(\frac{\sqrt{2m_*}}{\hbar}\nu z\right)\right]\nonumber
\end{eqnarray}
\begin{eqnarray}
&=&\frac1{2\pi}\frac{2m_*}{\hbar^2}2\nu\frac{\partial\nu}{\partial z}\frac1{\left[1+\frac{\sqrt{2m_*}}{\hbar}\nu zK_1\left(\frac{\sqrt{2m_*}}{\hbar}\nu z\right)\right]}\nonumber\\
&+&\frac1{2\pi}\frac{(2m_*)^2}{\hbar^4}\nu^3 z\frac{\left(z\frac{\partial\nu}{\partial z}+\nu\right)}{\left[1+\frac{\sqrt{2m_*}}{\hbar}\nu z K_1\left(\frac{\sqrt{2m_*}}{\hbar}\nu z\right)\right]^2}K_0\left(\frac{\sqrt{2m_*}}{\hbar}\nu z\right)\nonumber\\
&=&\frac1{2\pi}\frac{2m_*}{\hbar^2}\nu\left[-\frac{\nu}{2 z}-\frac1{8}\frac{2m_*}{\hbar^2}\nu^3z\ln\left(\frac{\sqrt{2m_*}}{2\hbar}\nu z\right)\right]\left[1-\frac1{4}\frac{2m_*}{\hbar^2}\nu^2z^2\ln\left(\frac{\sqrt{2m_*}}{2\hbar}\nu z\right)\right]\nonumber\\
&+&\frac1{8\pi}\frac{(2m_*)^2}{\hbar^4}\nu^3 z\left[\frac{\nu}{2}-\frac1{8}\frac{2m_*}{\hbar^2}\nu^3z^2\ln\left(\frac{\sqrt{2m_*}}{2\hbar}\nu z\right)\right]\left[1-\frac1{2}\frac{2m_*}{\hbar^2}\nu^2z^2\ln\left(\frac{\sqrt{2m_*}}{2\hbar}\nu z\right)\right]\nonumber\\
&\ &\qquad\qquad\qquad\qquad\times\left[-\ln\left(\frac{\sqrt{2m_*}}{2\hbar}\nu z\right)-\gamma\right]\nonumber\\
&=&-\frac1{2\pi}\frac{2m_*}{\hbar^2}\frac{\nu^2}{2 z}+O\left(\frac{(2m_*)^2}{\hbar^4}\nu^4 z\right)+O\left(\frac{(2m_*)^2}{\hbar^4}\nu^4 z\ln\left(\frac{\sqrt{2m_*}}{2\hbar}\nu z\right)\right)+...\nonumber\\
\end{eqnarray}
In our computations this term appears in the following combination, which comes from $\int \eta_+\eta_-$ term, it is denoted by $(3_b)$ in the main text,
\begin{eqnarray}
&\ &-\frac{\hbar^2}{M}2A\partial_z A\frac{\pi\hbar}{2m_*}\partial_z|_\nu\left[\frac{\sqrt{2m_*}}{\nu}zK_1\left(\frac{\sqrt{2m_*}}{\hbar}\nu z\right)\right]=\nonumber\\
&=&\!\!-\frac{\hbar^2}{M}2A\partial_z A\frac{\pi\hbar}{2m_*}\Bigg[\frac{\sqrt{2m_*}}{\nu}K_1\left(\frac{\sqrt{2m_*}}{\hbar}\nu z\right)\nonumber\\
&\ & \quad \quad \quad \quad \quad \quad \quad \quad  \quad \quad \quad \quad \quad-\frac{2m_*}{\hbar}z\bigg[K_0\left(\frac{\sqrt{2m_*}}{\hbar}\nu z\right)
 +\frac{\hbar}{\sqrt{2m_*}\nu z}K_1\left(\frac{\sqrt{2m_*}}{\hbar}\nu z\right)\bigg]\Bigg]\nonumber\\
&=&\frac{\hbar^2}{M}\left({2\over A}\partial_z A\right) A^2\frac{\pi\hbar}{2m_*}\frac{2m_*}{\hbar}zK_0\left(\frac{\sqrt{2m_*}}{\hbar}\nu z\right)\nonumber\\
&\approx&O\left(\frac{m_*}{M}\nu^2\right)+O\left(\frac{m_*}{M}\nu^2\ln\left(\frac{\sqrt{2m_*}}{2\hbar}\nu z\right)\right)+O\left(\frac{m_*}{M}\frac{2m_*}{\hbar^2}\nu^4z^2\right)+O\left(\frac{m_*}{M}\frac{2m_*}{\hbar^2}\nu^4z^2\ln\left(\frac{\sqrt{2m_*}}{2\hbar}\nu z\right)\right)\nonumber
\end{eqnarray}

The most complicated combinations come from the second derivative of the normalization constant $A$ with respect to $z$.  This  straight forward, yet long  computation can be simplified into nicer blocks by using some of the relations we have found. Let us recall the first derivative we have, 
\begin{eqnarray}
\partial_z A&=&\frac1{\sqrt{2\pi}}\frac{\sqrt{2m_*}}{\hbar}\frac{\partial\nu}{\partial z}\frac1{\sqrt{1+\frac{\sqrt{2m_*}}{\hbar}\nu z  K_1\left(\frac{\sqrt{2m_*}}{\hbar}\nu z\right)}}\nonumber\\
&+&\frac1{2\sqrt{2\pi}}\frac{\sqrt{2m_*}}{\hbar}\nu\left[\frac{\frac{2m_*}{\hbar^2}\nu z\left(z\frac{\partial\nu}{\partial z}+\nu\right)}{\left(1+\frac{\sqrt{2m_*}}{\hbar}\nu zK_1\left(\frac{\sqrt{2m_*}}{\hbar}\nu z\right)\right)^\frac{3}{2}}\right]K_0\left(\frac{\sqrt{2m_*}}{\hbar}\nu z\right)\nonumber
.\end{eqnarray}
We compute the second derivative, after tedious calculations and some simplifications, we arrive at 
\begin{eqnarray}
\partial^2_z A&=&\frac1{\sqrt{2\pi}}\frac{\sqrt{2m_*}}{\hbar}\frac{\partial^2\nu}{\partial z^2}\frac1{\sqrt{1+\frac{\sqrt{2m_*}}{\hbar}\nu z K_1\left(\frac{\sqrt{2m_*}}{\hbar}\nu z\right)}}\nonumber\\
&+&\frac1{2\sqrt{2\pi}}\frac{\sqrt{2m_*}}{\hbar}\frac{\partial\nu}{\partial z}\frac{\frac{2m_*}{\hbar^2}\nu z\left(z\frac{\partial\nu}{\partial z}+\nu\right)}{\left(1+\frac{\sqrt{2m_*}}{\hbar}\nu z K_1\left(\frac{\sqrt{2m_*}}{\hbar}\nu z\right)\right)^\frac{3}{2}}K_0\left(\frac{\sqrt{2m_*}}{\hbar}\nu z\right)\nonumber\\
&+&\frac1{2\sqrt{2\pi}}\frac{\sqrt{2m_*}}{\hbar}\frac{\partial\nu}{\partial z}\frac{\frac{2m_*}{\hbar^2}\nu z\left( z\frac{\partial\nu}{\partial z}+\nu\right)}{\left(1+\frac{\sqrt{2m_*}}{\hbar}\nu z K_1\left(\frac{\sqrt{2m_*}}{\hbar}\nu z\right)\right)^\frac{3}{2}}\nonumber\\
&+&\frac1{2\sqrt{2\pi}}\frac{\sqrt{2m_*}}{\hbar}\nu K_0\left(\frac{\sqrt{2m_*}}{\hbar}\nu z\right)\frac{2m_*}{\hbar^2}\frac{\left[2z^2\left(\frac{\partial\nu}{\partial z}\right)^2+\nu z^2\frac{\partial^2\nu}{\partial z^2}+5\nu  z\frac{\partial\nu}{\partial z}+\nu^2\right]}{\left(1+\frac{\sqrt{2m_*}}{\hbar}\nu zK_1\left(\frac{\sqrt{2m_*}}{\hbar}\nu z\right)\right)^\frac{3}{2}}\nonumber\\
&+&\frac{3}{4\sqrt{2\pi}}\frac{\sqrt{2m_*}}{\hbar}\nu \left(K_0\left(\frac{\sqrt{2m_*}}{\hbar}\nu z\right)\right)^2\frac{2m_*}{\hbar^2}\frac{\frac{2m_*}{\hbar^2}\nu^2z^2\left( z\frac{\partial\nu}{\partial z}+\nu\right)^2}{\left(1+\frac{\sqrt{2m_*}}{\hbar}\nu zK_1\left(\frac{\sqrt{2m_*}}{\hbar}\nu z\right)\right)^\frac{5}{2}}.\nonumber
\end{eqnarray}
What we need is the following combination,
\begin{eqnarray}
&\ &\frac1{A}\partial_z^2 A=\frac1{\nu}\frac{\partial^2\nu}{\partial z^2}+\frac1{2}\frac{\partial\nu}{\partial z}\frac{2m_*}{\hbar^2}\left[\frac{z\left(z\frac{\partial\nu}{\partial z}+\nu\right)}{\left(1+\frac{\sqrt{2m_*}}{\hbar}\nu zK_1\left(\frac{\sqrt{2m_*}}{\hbar}\nu z\right)\right)}\right]K_0\left(\frac{\sqrt{2m_*}}{\hbar}\nu z\right)\nonumber\\
&+&\frac1{2}\frac{\partial\nu}{\partial z}\frac{2m_*}{\hbar^2}\left[\frac{z\left(z\frac{\partial\nu}{\partial z}+\nu\right)}{\left(1+\frac{\sqrt{2m_*}}{\hbar}\nu zK_1\left(\frac{\sqrt{2m_*}}{\hbar}\nu z\right)\right)}\right]\nonumber\\
&+&\frac1{2} K_0\left(\frac{\sqrt{2m_*}}{\hbar}\nu z\right)\frac{2m_*}{\hbar^2}\left[\frac{2z^2\left(\frac{\partial\nu}{\partial z}\right)^2+5\nu z\frac{\partial\nu}{\partial z}+\nu z^2\frac{\partial^2\nu}{\partial z^2}+\nu^2}{\left(1+\frac{\sqrt{2m_*}}{\hbar}\nu zK_1\left(\frac{\sqrt{2m_*}}{\hbar}\nu z\right)\right)}\right]\nonumber\\
&+&\frac{3}{4}\frac{2m_*}{\hbar^2}\left[\frac{\frac{2m_*}{\hbar^2}\nu^2z^2\left(z\frac{\partial\nu}{\partial z}+\nu\right)^2}{\left(1+\frac{\sqrt{2m_*}}{\hbar}\nu zK_1\left(\frac{\sqrt{2m_*}}{\hbar}\nu z\right)\right)^2}\right]\left(K_0\left(\frac{\sqrt{2m_*}}{\hbar}\nu z\right)\right)^2
.\end{eqnarray}
In our approach, we need the small $z$ expansion of ${1\over A} \nabla_z^2 A$, which in cylindrical coordinates, can be expressed with the first and second $z$ derivatives of $A$. 
As one may appreciate the expression above is fairly complicated, its exact expansion requires care and patience, since we are not actually computing the second order correction, we will only point out the types of terms we  encounter in this expansion, with some undetermined coefficients in front.
As a consequence, the Laplacian part  of normalization is found to be,
\begin{equation}
-\frac{\hbar^2}{M}\frac1{A}\nabla_z^2 A\approx-\frac{\hbar^2}{M}\frac1{4z^2}+c_1\frac{m_*}{M}\nu^2\ln\left(\frac{\sqrt{2m_*}}{\hbar}\nu z\right)+c_2\frac{m_*}{M}\nu^2+c_3\frac{m_*}{M}\frac{2m_*}{\hbar^2}\nu^4z^2\left(\ln\left(\frac{\sqrt{2m_*}}{\hbar}\nu z\right)\right)^2 +...\label{kineticA}
,\end{equation}
where $c_1,c_2,c_3$ are constant that can be found explicitly, moreover $m_*$ {\it should be replaced with $m$ to this order of accuracy}.
These are the results that we use in the main text. In the subsequent Appendix to illustrate the consistency of our approximations as well as proposing a scheme to compute higher order corrections we evaluate the expectation values of some of the next order terms within first order perturbation theory. This of course is an asymptotic expansion and hopefully presents a reliable description of  the dynamics.

\section{Appendix-Perturbative corrections}

In order to verify the consistency of our approximations we should first show that the expectation value of $z$ is much smaller than the spread of the light particles wave functions, which is characterized by 
$\zeta_0={\hbar\over \sqrt{2m_*}\epsilon}$. This is related to $z_0$ that we introduce previously, one can see easily  that $z_0={2m_*\over M} \zeta_0$. 
For this, we need to normalize our wave function, written in terms of the variable $z$;
\begin{equation}
\Psi(z)=C\frac{2^\beta z^\beta}{(z_0(1+2\beta))^\beta}e^{-\frac{z}{z_0 (1+2\beta)}}
,\end{equation}
it requires computing the integral below, hence we find the normalization, 
\begin{equation}
C^2 2\pi\frac{2^{2\beta} }{(z_0(1+2\beta))^{2\beta}}\int_0^\infty dz z^{2\beta+1}e^{-\frac{2z}{z_0 (1+2\beta)}}=1\qquad {\rm and}\qquad C=\sqrt{\frac{2}{\pi\Gamma(2\beta+2)z_0^2(1+2\beta)^2}}.
\end{equation}
Next we look at the expectation value of  $z$ (within our approximtion) to check the consistency of our assumption
\begin{eqnarray}
<z>&=&\frac{2^{2\beta+2}}{(z_0(1+2\beta))^{2\beta+2}\Gamma(2\beta+2)}\int_0^\infty dz z^{2\beta+2}e^{-\frac{2z}{z_0(1+2\beta)}}=\frac1{2}\frac{\Gamma(2\beta+3)}{\Gamma(2\beta+2)}z_0(1+2\beta)\nonumber\\
&=&\frac{e^\gamma}{2}\frac{\Gamma(2\beta+3)}{\Gamma(2\beta+2)}(1+2\beta)\frac{\hbar}{\sqrt{2m_*}\epsilon}\left(\frac{m_*}{M}\right)
.\end{eqnarray}
Let us emphasize that this is a key result for the consistency of our approximations.

It is also instructive to calculate next order corrections to the binding energy of the heavy-light system as a function of the distance between the heavy centers.
\begin{equation}
-\nu^2 \approx-\frac{\hbar}{\sqrt{2m_*}}\frac{2\epsilon}{e^\gamma z}+\frac{\epsilon^2}{2e^{2\gamma}}\ln\left(\frac{\sqrt{2m_*}}{2\hbar}\epsilon ze^\gamma\right)+...\nonumber\\
.\end{equation}
The first term  is the original potential part that we used in the effective description of the heavy system,  so let us  look at the expectation value of the second term as a perturbation on our solution:
\begin{eqnarray}
\bigg<\frac{\epsilon^2}{2e^{2\gamma}}\ln\left(\frac{\sqrt{2m_*}}{2\hbar}\epsilon ze^\gamma\right)\bigg>&=&\frac{\epsilon^2}{2e^{2\gamma}}\frac{2^{2\beta+2}}{(z_0(1+2\beta))^{2\beta+2}\Gamma(2\beta+2)}\int_0^\infty dz z^{2\beta+1}e^{-\frac{2z}{z_0(1+2\beta)}}\ln\left(\frac{\sqrt{2m_*}}{2\hbar}\epsilon ze^\gamma\right)\nonumber\\
&=&\frac{\epsilon^2}{2e^{2\gamma}}\bigg(\psi(2\beta+2)-\ln\left(\frac{M}{2m_*}\right)+\ln\left(\frac{(1+2\beta)e^{2\gamma}}{8}\right)\bigg)\nonumber\\
&\approx& -\frac{\epsilon^2}{2e^{2\gamma}}\ln\left(\frac{M}{m_*}\right)
.\end{eqnarray}
 We can easily see that this expectation value  is of lower order compared to  the leading order energy of the heavy particle. We emphasize that the constant terms are ambigous as long as we keep to this order, they can be fixed if the next order terms ($z$ without the logarithms) are also taken into account.
An interesting estimate will be to calculate the expectation values of the correction terms resulting from the heavy particle relative coordinate kinetic energy term (Laplacian with respect to $z$) operating on the normalization constant $A$, the expansion of which is given in the previous Appendix in equation (\ref{kineticA}). Since we have not found the precise constant in front, we compute the expectation values of each basic part separately without the constants in front:
\begin{eqnarray}
(a)&=&\frac{2m_*}{M}\nu^2\approx\frac{2m_*}{M}\frac{\hbar}{\sqrt{2m_*}}\frac{2\epsilon}{e^\gamma z}-\frac{2m_*}{M}\frac{\epsilon^2}{2e^{2\gamma}}\ln\left(\frac{\sqrt{2m_*}}{2\hbar}\epsilon ze^\gamma\right)+...\nonumber\\
(b)&=&\frac{2m_*}{M}\nu^2\ln\left(\frac{\sqrt{2m_*}}{2\hbar}\nu z e^\gamma\right)\approx \frac1{2} \frac{2m_*}{M}\frac{\hbar}{\sqrt{2m_*}}\frac{2\epsilon}{e^\gamma z}\ln\left(\frac{\sqrt{2m_*}}{2\hbar}\epsilon ze^\gamma\right)-\frac1{4e^{2\gamma}}\frac{2m_*}{M}\epsilon^2\ln^2\left(\frac{\sqrt{2m_*}}{2\hbar}\epsilon z e^\gamma\right)+...\nonumber\\
(c)&=&\frac{2m_*}{M}\frac{2m_*}{\hbar^2}\nu^4z^2\ln^2\left(\frac{\sqrt{2m_*}}{2\hbar}\nu ze^\gamma\right)\approx\frac{2m_*}{M}\frac{\epsilon^2}{e^{2\gamma}}\ln^2\left(\frac{\sqrt{2m_*}}{2\hbar}\epsilon ze^\gamma\right)+...
.\end{eqnarray}
We first look at the expectation value of $ (a)$  the log-term has already been calculated above so let us focus on  the expectation of the first term  of $(a)$, that we denote by a subscript as $(a)_1$, 
\begin{eqnarray}
<(a)_1>&=&\bigg<\frac{2m_*}{M}\frac{\hbar}{\sqrt{2m_*}}\frac{2\epsilon}{e^\gamma z}\bigg>=\frac{2m_*}{M}\frac{\hbar}{\sqrt{2m_*}}\frac{2\epsilon}{e^\gamma}\frac{2^{2\beta+2}}{(z_0(1+2\beta))^{2\beta+2}\Gamma(2\beta+2)}\int_0^\infty dz z^{2\beta}e^{-\frac{2z}{z_0(1+2\beta)}}\nonumber\\
&=&\frac{2m_*}{M}\frac{\hbar}{\sqrt{2m_*}}\frac{4\epsilon}{e^\gamma}\frac{\Gamma(2\beta+1)}{\Gamma(2\beta+2)z_0(1+2\beta)}=\frac{8}{e^{2\gamma}}\frac{\epsilon^2}{(1+2\beta)^2}\nonumber
,\end{eqnarray}
which is negligible to this order.
In a similar way, we find the expectation value of the second term $(b)$  
\begin{eqnarray}
<(b)>&=&\bigg<\frac{2m_*}{M}\frac{\hbar}{\sqrt{2m_*}}\frac{\epsilon}{e^\gamma z}\ln\left(\frac{\sqrt{2m_*}}{2\hbar}\epsilon ze^\gamma\right)\bigg>\nonumber\\
&=&\frac{2m_*}{M}\frac{\hbar}{\sqrt{2m_*}}\frac{\epsilon}{e^\gamma}\frac{2^{2\beta+2}}{(z_0(1+2\beta))^{2\beta+2}\Gamma(2\beta+2)}\int_0^\infty dz z^{2\beta}e^{-\frac{2z}{z_0(1+2\beta)}}\ln\left(\frac{\sqrt{2m_*}}{2\hbar}\epsilon ze^\gamma\right)\nonumber\\
&=&\frac{2m_*}{M}\frac{\hbar}{\sqrt{2m_*}}\frac{2\epsilon}{e^\gamma z_0(1+2\beta)}\frac{\Gamma(2\beta+1)}{\Gamma(2\beta+2)}\bigg(\psi(2\beta+1)-\ln\left(\frac{M}{2m_*}\right)+\ln\left(\frac{(1+2\beta)e^{2\gamma}}{8}\right)\bigg)\nonumber\\
&=&\frac{4}{e^{2\gamma}}\frac{\epsilon^2}{(1+2\beta)^2}\bigg(\psi(2\beta+1)-\ln\left(\frac{M}{2m_*}\right)+\ln\left(\frac{(1+2\beta)e^{2\gamma}}{8}\right)\bigg)\nonumber\\
&\approx&-\frac{4\epsilon^2}{e^{2\gamma}}{1\over (1+2\beta)^2}\ln\left(\frac{M}{m_*}\right)
\end{eqnarray}
The final expectation value is that of  $(c)$ part, it must be negligible due to $m/M$ term and no inverse powers of $z$ appearing in it, 
\begin{eqnarray}
<(c)>&=&\bigg<\frac{2m_*}{M}\frac{\epsilon^2}{e^{2\gamma}}\ln^2\left(\frac{\sqrt{2m_*}}{2\hbar}\epsilon ze^\gamma\right)\bigg>\nonumber\\
&=&\frac{2m_*}{M}\frac{\epsilon^2}{e^{2\gamma}}\frac{2^{2\beta+2}}{(z_0(1+2\beta))^{2\beta+2}\Gamma(2\beta+2)}\int_0^\infty dz z^{2\beta+1}e^{-\frac{2z}{z_0(1+2\beta)}}\ln^2\left(\frac{\sqrt{2m_*}}{2\hbar}\epsilon ze^\gamma\right)\nonumber\\
&=&\frac{2m_*}{M}\frac{\epsilon^2}{e^{2\gamma}}\Bigg[\bigg(\psi(2\beta+2)-\ln\left(\frac{M}{2m_*}\right)+\ln\left(\frac{(1+2\beta)e^{2\gamma}}{8}\right)\bigg)^2+\zeta(2,2\beta+2)\Bigg]
,\end{eqnarray}
which is {\it negligible} to the order that we are interested in, as expected.
To gain more insight, we can go one step further and calculate  some of the expectation values resulting  from the heavy particle kinetic energy partially operating on  the light particle wave function and generating a mixed gradient term, after our simplifications, this residual term is found to be,
\begin{eqnarray}
\frac{2m_*}{M}\nu^2 z\frac{\partial \Psi}{\partial z}\ln\left(\frac{\sqrt{2m_*}}{2\hbar}\nu ze^\gamma\right)&\approx&\frac{m_*}{M}\frac{\hbar}{\sqrt{2m_*}}\frac{2\epsilon}{e^\gamma}\frac{\partial \Psi}{\partial z}\ln\left(\frac{\sqrt{2m_*}}{2\hbar}\epsilon ze^\gamma\right)\nonumber\\
&-&\frac{2m_*}{M}\frac{\epsilon^2}{4e^{2\gamma}}z\frac{\partial\Psi}{\partial z}\ln^2\left(\frac{\sqrt{2m_*}}{2\hbar}\epsilon ze^\gamma\right)+...\nonumber
\end{eqnarray}
We compute the expectation value of this expression (to properly identify the corrections within first order perturbation theory we must  symmetrize this expression, thus we get a term coming from the anticommutator of ${\partial \over \partial z}$ and the log expression, we ignore this subtlety for now since we are not aiming for an exact computation), let us start with the first part,
\begin{eqnarray}
&\ &\frac{2m_*}{M}\frac{\hbar}{\sqrt{2m_*}}\frac{\epsilon}{e^\gamma}2\pi\int_0^\infty dz z\Psi\frac{\partial\Psi}{\partial z}\ln\left(\frac{\sqrt{2m_*}}{2\hbar}\epsilon ze^\gamma\right)=\frac{m_*}{M}\frac{\hbar}{\sqrt{2m_*}}\frac{\epsilon}{e^\gamma}2\pi\int_0^\infty dz z\frac{\partial\Psi^2}{\partial z}\ln\left(\frac{\sqrt{2m_*}}{2\hbar}\epsilon ze^\gamma\right)\nonumber\\
&=&\frac{m_*}{M}\frac{\hbar}{\sqrt{2m_*}}\frac{\epsilon}{e^\gamma}2\pi\left[\Psi^2 z\ln \left(\frac{\sqrt{2m_*}}{2\hbar}\epsilon ze^\gamma\right)\big|_0^\infty-\int_0^\infty dz\Psi^2\ln\left(\frac{\sqrt{2m_*}}{2\hbar}\epsilon ze^\gamma\right)-\int_0^\infty dz \Psi^2 \right]\nonumber\\
&=&-\frac{m_*}{M}\frac{\hbar}{\sqrt{2m_*}}\frac{\epsilon}{e^\gamma}\left[<{1\over z}>+2\pi\int_0^\infty dz\Psi^2\ln\left(\frac{\sqrt{2m_*}}{2\hbar}\epsilon ze^\gamma\right)\right]\nonumber\\
&\approx& \frac{2\epsilon^2}{e^{2\gamma}(1+2\beta)^2}\ln\left(\frac{M}{m_*}\right).
\end{eqnarray}
The second part of the expansion should be small, let us see this again in our explicit computation (within first order perturbation theory),
\begin{eqnarray}
&\ &-\frac{2m_*}{M}\frac{\epsilon^2}{8e^{2\gamma}}2\pi\int_0^\infty dz z^2\frac{\partial \Psi^2}{\partial z}\ln^2\left(\frac{\sqrt{2m_*}}{2\hbar}\epsilon z e^\gamma\right)\nonumber\\
&=&-\frac{2m_*}{M}\frac{\epsilon^2}{8e^{2\gamma}}2\pi\Bigg[z^2\Psi^2\ln^2\left(\frac{\sqrt{2m_*}}{2\hbar}\epsilon ze^\gamma\right)\big|_0^\infty-2\int_0^\infty dz z \Psi^2\ln^2\left(\frac{\sqrt{2m_*}}{2\hbar}\epsilon ze^\gamma\right)\nonumber\\
&\ &-2\int_0^\infty dz z\Psi^2\ln\left(\frac{\sqrt{2m_*}}{2\hbar}\epsilon ze^\gamma\right)\Bigg]\nonumber\\
&=&\frac{2m_*}{M}\bigg<\frac{\epsilon^2}{4e^{2\gamma}}\ln\left(\frac{\sqrt{2m_*}}{2\hbar}\epsilon ze^\gamma\right)\bigg>+\frac{2m_*}{M}\bigg<\frac{\epsilon^2}{4e^{2\gamma}}\ln^2\left(\frac{\sqrt{2m_*}}{2\hbar}\epsilon ze^\gamma\right)\bigg>,
\end{eqnarray}
all of which are computed above and clearly of smaller order.
This completes our short digression on calculating higher order corrections, as we emphasize {\it in principle } it is possible to compute all these corrections in first order perturbation theory, nevertheless it requires precise  expansions of all the terms to first order (in terms of $z$) which we have not done. If we intend to go beyond the first order terms in the expansions and evaluate their contributions, the possibility of second order perturbation of the first oder terms becoming equally important shoud be discussed. We leave these questions to the future.


\begin{thebibliography}{99}
\bibitem{Born-Opp}M. Born and J. R. Oppenheimer  "Zur Quantentheorie der Molekeln" [On the Quantum Theory of Molecules]. Annalen der Physik  389 (20) (1927), pg. 457 - 484. 
\bibitem{landau} L. D. Landau and E. M. Lifshitz, {\it Quantum Mechanics: Nonrelativistic Theory}, Butterworth-Heinemann; 3rd edition (1981).
\bibitem{bethe-jackiw} H. Bethe and R. Jackiw, {\it Intermediate Quantum Mechanics},  W. A. Benjamin, Inc.; 2nd edition (1973).
\bibitem{weinbergQM} S. Weinberg, {\it Lectures on Quantum Mechanics}, Cambridge University Press; 2nd edition (2015).
\bibitem{Spohn}G. Panati, H. Spohn and S. Teufel, {\it The Time-Dependent Born-Oppenheimer Approximation}, ESAIM: Math. Modelling and Numerical Analysis 41, 297-314 (2007)
\bibitem{Hagedorn-Joye} G. A. Hagedorn and T. Joye, {\it Mathematical Analysis of Born-Oppenheimer Approximations}, Spectral
theory and Mathematical Physics: A Festschrift in Honor of Barry Simon's 60th Birthday :Quantum Field Theory, Statistical Mechanics and Non-relativistic Quantum Systems, edited by F. Gesztesy et. al. , AMS publications (2007).
\bibitem{Jecko-5} T. Jecko, {\it On the mathematical treatment of the Born-Oppenheimer approximation}, Jour. of Math. Phys. 55, 053504 (2014)

\bibitem{Thorn} C. Thorn, {\it Quark confinement in the infinite momentum frame} Phys. Rev. D19, pg 639 (1979)
\bibitem{Beg} M. A. B. Beg and R. C. Furlong, $\lambda \phi^4$ {\it theory in the nonrelativistic limit},  Phys.  Rev.  D31, pg 1370 (1985)
\bibitem{Jackiw} R. Jackiw "{\it Delta-function potentials in two and three dimensional Quantum mechanics}" M.A.B.beg memorial volume, A. Ali and P. Hoodbhoy Editors, World Scientific Singapore, (1991).
\bibitem{Perez} J. Fernando Perez and F. A. B. Coutinho, {\it Schr\"odinger equation in two dimensions for a zero range potential and a uniform magnetic field: An exactly solvable model } Amer. Jour. Phys. Vol. 59, pg 52 (1991) 
 \bibitem{Mead} L. R. Mead and Godines, {\it An analytical example of renormalization in quantum mechanics} Amer. Jour. Phys. Vol. 59 pg 935 (1991).
\bibitem{Tarrach} P. Gosdzinsky and R. Tarrach {\it Learning quantum field theory from elementary quantum mechanics }Amer. Jour. Phys. Vol. 59, pg. 70 (1991)
\bibitem{Albeverio} S. Albeverio, F. Gesztesy, R. Hoegh-Krohn and H. Holden, {\it Solvable Models in Quantum Mechanics} 2nd Edition, AMS Chelsea Publishing, Providence, Rhode Island (2004).
\bibitem{Mitra} I. Mitra, Dasgupta and B. Dutta-Roy {\it Regularization and renormalization in scattering from Dirac delta potentials}, Amer. Jour. Phys. Am. J. Phys. Vol. 66, pg. 1101 (1998)
\bibitem{seiler-2} R. Seiler, {\it Does the Born-Oppenheimer Approximation Work?}, Helvetica Physica Acta, 46 (1973) 230.
\bibitem{combes-seiler}J. M. Combes, P. Duclos and R. Seiler, {\it  The  Born-Oppenheimer Approximation}, Wightman, Velo (Eds.), Rigorous Atomic and Molecular Physics Proceedings, 1980, Plenum, New York (1981), pp. 185
\bibitem{hunziker}W. Hunziker, {\it Distortion Analyticity and Molecular Resonance Curves}, Ann. Inst. H.
Poincare Sect. A. 45 (1986), 339.
\bibitem{hagedorn4}G. A. Hagedorn, {\it Classification and Normal Forms for Quantum Eigenvalue Crossings} Asterisque 210 (1993), 115-134.
\bibitem{hagedorn5}G. A. Hagedorn, {\it Molecular Propagation through Electron Energy Level Crossings}, Memoirs
Amer. Math. Soc. 111 (536) (1994), 1-130.
\bibitem{hagedorn6}G. A. Hagedorn, {\it Classification and Normal Forms for Avoided Crossings of Quantum Mechanical
Energy Levels}, Jour. Phys. A. 31 (1998), 369-383.

\bibitem{hagedorn1} G. A. Hagedorn, {\it High Order Corrections to the Time-Independent Born-Oppenheimer Approximation
I: Smooth Potentials}, Ann. Inst. H. Poincare Sect. A. 47 (1987), 1-16.
\bibitem{hagedorn2}G. A. Hagedorn, {\it  High Order Corrections to the Time-Independent Born-Oppenheimer Approximation
II: Diatomic Coulomb Systems}, Commun. Math. Phys. 116 (1988), 23-44.
\bibitem{seiler-klein}  M. Klein, A. Martinez, R. Seiler, X. Wang, {\it On the Born-Oppenheimer Expansion for Polyatomic
Molecules}, Commun. Math. Phys. 143 (1992), 607-639.

\bibitem{weigert} S. Weigert and R. G. Littlejohn, {\it Diagonalization of multicomponent wave equations with a Born-Oppenheimer example}, Phys. Rev. A, Vol. 45 (1993) 3506.

\bibitem{dutta-roy} G. Gangopadhyay and B. Dutta-Roy, {\it The Born-Oppenheimer Approximation: the Toy Model}, American Jour. Phys. 72 (2004) pg 389
.
\bibitem{haci-turgut} H. Akbas and O. T. Turgut, {\it Born-Oppenheimer Approximation for a singular system}, Arxiv-1602-02811, submitted for publication.



\bibitem{gradsh} I.S. Gradshtein and I.M. Ryzhik, {\it Table of Integrals, Series, and Products}, Academic Press (2007).
\end{thebibliography}
\end{document}